\documentclass[trackchanges, twocolumn]{aastex701}

\begin{document}

\correspondingauthor{Vaibhav Pant}
\email{vpant@iitd.ac.in}

\title{CME Kinematics with Optical Flow: Multi-Coronagraph Insights into Internal Velocity Dispersion}

\author[orcid=0009-0005-9842-709X,gname='Pritam',sname='Das']{Pritam Das}
\affiliation{Aryabhatta Research Institute of Observational Sciences, Nainital-263001, India}
\affiliation{Department of Applied Physics, Mahatma Jyotiba Phule Rohilkhand University, Bareilly-243006, India}
\email{pritamd9818@gmail.com}  

\author[orcid=0000-0002-6954-2276,gname=Vaibhav, sname='Pant']{Vaibhav Pant}
\affiliation{Department of Physics, Indian Institute of Technology Delhi, Delhi-110016, India}
\email{vpant@iitd.ac.in}

\author[gname=Neeraj,sname=Rathore]{Neeraj Rathore}
\affiliation{Department of Applied Physics/Physics, Bareilly College, Bareilly-243001, India}
\email{rathorenjs@gmail.com}

\begin{abstract}

Coronal mass ejections (CMEs) exhibit poorly understood internal velocity distributions. In this work, we introduce an optical flow-based method, DOFCAT (Dense Optical Flow CME Analysis Tool), applied to high-resolution data from ASPIICS/Proba-3 and METIS/Solar Orbiter to examine the internal velocity dispersion of different substructures within CMEs in the Middle Corona. We also introduce a Gaussian-tapered Fourier (GTF) filter that suppresses brightness artifacts in running-difference images from the ASPIICS wideband channel. We then apply DOFCAT to four CME events observed across ASPIICS and METIS, extracting their internal velocity information. Our results show that the velocity profiles of both the CME front and core change with position angle and are non-uniform. For impulsive CMEs, the height offset between the front and core begins to increase following the impulsive acceleration phase, while CMEs without a clear impulsive phase show a more gradual evolution of the height offset. We find that this increase in height difference occurs rapidly between 2.3--3.0\,R$_\odot$. Furthermore, the front shows a higher velocity spread than the core, with impulsiveness evident in the CME leading half. We further compute the radial self-similarity coefficient separately for the front and the core, finding that the front expands non-self-similarly initially and tends toward self-similarity at larger heights, while the core remains relatively self-similar. These findings confirm that CMEs contain complex internal flows and layered substructures within the traditional three-part morphology. Our results demonstrate the utility of DOFCAT on high-resolution coronagraphs for probing CME dynamics and provide new insights into CME initiation and evolution.

\end{abstract}



\section{Introduction}
\label{Introduction}

Coronal Mass Ejections (CMEs) are spectacular solar eruptions that have fascinated heliophysicists and space weather researchers for a long time. Since their discovery in the early 1970s \citep{Hansen1971}, CMEs have been recognised as major drivers of space weather. They release large amounts of magnetized plasma from the solar atmosphere into the heliosphere, often causing strong geomagnetic storms that can affect both space-based and ground-based technology \citep{Webb2012, Temmer2021}.

Over the past decades, white-light coronagraphs have revealed a range of CME morphologies. Among these, the so-called ``three-part structure'' CME \citep{Illing1985} consisting of a bright leading edge (LE), a dark cavity, and a bright core has received particular attention for its connection to erupting flux ropes \citep{Vourlidas2013, Song2023, Song2025}. Such structured CMEs are especially valuable for studying internal dynamics, as their substructures can often be identified and tracked independently. Furthermore, the traditional concept of the three-part CME structure has been increasingly challenged with the advent of high-resolution imaging instruments, which reveal complex internal morphologies and additional substructures beyond the classical leading edge, cavity, and core \citep{Wood2021, Guo2025}. However, detecting and investigating the kinematics of these features in the middle corona (1.5–6.0~R$_\odot$) \citep{West2023}  has remained challenging due to instrumental limitations.

Much of our understanding of CME kinematics comes from global measurements such as CME LE's height-time profiles and average speeds. These often show a three-phase evolution: a slow rise, followed by impulsive acceleration, and finally a propagation phase \citep{Chen2003, Zhang2004}. \citet{Pant2021} classified CME velocities from the manual CDAW catalog into three categories— slow, intermediate, and fast, based on the average solar wind speed of 400 $\mathrm{km\ s^{-1}}$. While this framework has proven useful, it provides only a partial view of CME kinematics, considering only the average LE speed while not accounting for the internal variations within the CME. Crucial details about the internal motion, particularly the relative velocities of the front and core, remain elusive, yet these are precisely the regions where the main acceleration occurs. Moreover, growing evidence suggests that different parts of the CME may have differential volume expansion, indicating internal velocity dispersion \citep{Majumdar2022}. The internal velocity structure of CMEs can hold important clues about the underlying eruption mechanisms and thermodynamic evolution \citep{Bemporad2018}.

A growing body of work has begun to address these internal dynamics. Several studies have noted that the CME leading edge often travels faster than the core \citep{Maričić2009}. This velocity difference leads to significant radial expansion, which governs the thermodynamic state of the plasma; recent diagnostics, for instance, reveal that such expanding CMEs often maintain nearly isothermal profiles, implying internal heating processes \citep{Sheoran2023}. More recent work, such as \citet{Majumdar2024}, has quantified front–core velocity dispersion, showing that the height difference between the front and the core begins to increase at a critical height $h_{\mathrm{c}}$ between 1.4 and 1.8~R$_\odot$. This reinforces the idea that CMEs are internally dynamic structures rather than rigidly expanding shells.

A number of existing CME tracking techniques also face limitations when probing internal kinematics. The CDAW catalog, for instance, is based on manual visual identification and height-time measurements of CME LE from SOHO coronagraph images \citep{Yashiro2004}. While it has been instrumental in building a comprehensive database, its manual nature introduces some subjectivity and focuses primarily on the LE kinematics. The widely used GCS (Graduated Cylindrical Shell) model approximates the CME shape using a croissant-shaped flux rope shell and fits it to multi-viewpoint coronagraph images \citep{Thernisien2009}. However, this model primarily tracks the global outline of the CME, limiting its ability to capture differential motion within the CME.

Automated methods like CACTus (Computer Aided CME Tracking) apply a linear Hough transform to detect bright ridges in height-time plots of coronagraph images, essentially assuming a linear velocity profile \citep{Robbrecht2004, Pant2016}. CIISCO improves upon this using a parabolic Hough transform to identify curved trajectories \citep{Patel2021}, yet still presumes a specific acceleration profile that may not generalize to all CME events. Other approaches fit spline curves to height-time measurements \citep{Majumdar2022}, allowing for more flexibility but still depending critically on the visibility and manual selection of points along the CME front and core.

Beyond front-tracking approaches, cross-correlation based methods have been employed to estimate spatially resolved CME speeds. For example, \citet{Ying2019} derived radial velocity maps by correlating intensity profiles between successive coronagraph images; however, this approach is primarily sensitive to bulk radial translation and is less capable of capturing non-radial motions, internal deformation, or differential expansion within the CME body.

Collectively, these limitations highlight the need for alternative methods that directly measure velocity fields without relying on geometric assumptions or predefined trajectories. One promising approach is through optical flow techniques and computer vision algorithms originally developed to estimate motion between image frames. \citet{Colaninno2006} and \citet{Chen2024} applied optical flow to LASCO C2 images and demonstrated its utility in tracking the motion of CME substructures across the field of view.

The advent of next-generation coronagraphs has created new opportunities for advancing this line of inquiry. METIS aboard \textit{Solar Orbiter} \citep{Antonucci2020, Fineschi2020, DeLeo2023} and ASPIICS on Proba-3 \citep{Zhukov2025} now provide continuous imaging of the corona from 1.7 to 3.1~R$_\odot$ (at minimum perihelion) and 1.099 to 3.0~R$_\odot$ respectively, with high spatial resolution and cadence. These instruments are particularly well-suited for studying the early and middle phases of CME evolution, where most of the acceleration and internal restructuring takes place.

In this work, we build on these advances by presenting a robust optical flow–based method to study internal velocity dispersion in CMEs observed by ASPIICS and METIS. We analyse four structured CME events, examining how internal velocity profiles, particularly between the front and core, vary with height, position angle, and impulsiveness. This analysis offers insight into the coupling and decoupling of CME substructures, with implications for initiation models and thermodynamic evolution. Section~\ref{sec:data} describes the instruments and events studied, followed by image preprocessing and optical flow methodology in Section~\ref{sec:Methodology}. The results are presented in Section~\ref{sec:Results}, and concluded in Section~\ref{sec:Discussion}.

\section{Instrument and Data}
\label{sec:data}

\subsection{Data Source}
\label{sec:data_source}
This study utilises data from two high-resolution space-based coronagraphs observing the inner to middle corona: ASPIICS (Association of Spacecraft for Polarimetric and Imaging Investigation of the Corona of the Sun) onboard Proba-3 mission and METIS (Multi Element Telescope for Imaging and Spectroscopy)aboard ESA’s Solar Orbiter.

ASPIICS is the first coronagraph to operate via formation flying, using two spacecraft to achieve an artificial eclipse. This innovative configuration enables uninterrupted imaging of the solar corona from 1.099 to 3.0~R$\odot$ with high spatial resolution. We use Level 2 wideband images centred at 5510.6~\AA. Among the available exposure durations (0.1, 1, and 10 seconds) \citep{Zhukov2025b}, ASPIICS primarily uses the 1- and 10-second exposures to enhance contrast in the inner and middle corona, respectively. However, a micro-level misalignment between the external occulter and the telescope introduces brightness flickering, most prominently visible in running difference sequences.

We were able to mitigate this flickering in the 10-second exposure images (Section~\ref{sec:ImagePreProc}), but not as effectively in the 1-second channel, which limited its suitability for optical flow analysis. We therefore selected the 10-second channel, as it provided stable images suitable for tracking CME motion in the 1.6 to 3.0~R$\odot$ range.

METIS (Multi Element Telescope for Imaging and Spectroscopy), onboard Solar Orbiter, is a dual-channel coronagraph that simultaneously observes the corona in visible light (VL) from 5800 to 6400~\AA\ and the ultraviolet (UV) Lyman-$\alpha$ line at 1216~\AA\ \citep{Antonucci2020}. We use Level 2 data from the \href{https://metis.oato.inaf.it/metis_data_release_1.0.html}{METIS Data Release 1.0}. The VL images are used for optical flow analysis to extract velocity fields, while co-temporal UV images (if available) help identify associated prominence material, often corresponding to the CME core \citep{Illing1986}.

To further establish the spatial correspondence between the prominence and CME core, we additionally use images from the Full Sun Imager (FSI) onboard \textit{Solar Orbiter} \citep{Rochus2020}, particularly in the 304~\AA\ channel. These images provide context for the source region of the erupting material and enable cross-instrument feature association.

\subsection{Event Selection}

We selected four coronal mass ejection (CME) events observed between February 2020 (the launch of METIS) and August 2025. The selection was based on the following criteria:

\begin{enumerate}
    \item The CME must exhibit a clear three-part or multi-part structure in white-light images, with well-defined front and core components suitable for internal kinematic analysis.
    \item The selected CME events must qualify as limb CMEs (as per the perspective of the observing coronagraph), based on the definitions provided in \citet{Pant2021} and \citet{Gopalswamy2014}. This classification was verified using JHelioviewer \citep{Muller2017} and corroborated with multi-vantage point observations from the Atmospheric Imaging Assembly (AIA; \citealp{Lemen2012}) on the Solar Dynamics Observatory (SDO), the Solar Ultraviolet Imager (SUVI; \citealp{Darnel2022}) on Geostationary Operational Environmental Satellites (GOES), the Full Sun Imager (FSI; \citealp{Rochus2020}) on Solar Orbiter, and the Extreme Ultraviolet Imager (EUVI; \citealp{Howard2002, Kaiser2008}) on the Solar TErrestrial RElations Observatory (STEREO), where available.
    \item The available cadence for the relevant data must be better than two minutes to ensure sufficient temporal resolution for optical flow.This cadence requirement arises because Farnebäck optical flow estimation relies on the assumption of small, sub-pixel displacements between consecutive frames. Larger cadences (e.g., the 12-minute cadence LASCO C2) may violate this assumption and introduce temporal aliasing in the calculated velocity field, particularly for fast-moving CME substructures, which can traverse a large distance and undergo significant morphological changes within just a few minutes.
    \item Continuous data coverage must be available from the CME’s first appearance in the instrument’s field of view until the leading edge exits. Also, the inner FOV of the coronagraph must not be greater than $2.5~R_\odot$.
    
\end{enumerate}

Table~\ref{tab:event-summary} summarizes the details of data availability associated with each of the four CME events, including time intervals, observation channels, cadence, and complementary datasets.

\begin{deluxetable*}{lcccc}
\tablecaption{Summary of data availability associated with the four CME events.\label{tab:event-summary}}
\tablehead{
\colhead{Date} & 
\colhead{Time Interval (UT)} & 
\colhead{Observational Channel} & 
\colhead{Cadence (s)} & 
\colhead{Complementary Data}
}
\startdata
2025-07-16 & 16:35:08–17:32:07 & ASPIICS (WB), 10 s exposure & $\sim$30$^{\#}$ & SUVI 304~\AA \\
           & 17:32:07–17:33:39 &  & 90 &  \\
           & 17:33:39–18:02:07 &  & $\sim$30$^{\#}$ &  \\
\hline
2024-04-16 & 07:51:13–08:28:36 & METIS VL (tB) & 21 & METIS VL (pB) \\
           & 08:28:36–08:30:55 &  & 139 &  \\
           & 08:30:55–09:17:07 &  & $\sim$63$^{\#}$ &  \\
\hline
2024-03-29 & 20:29:31–21:17:16 & METIS VL (tB) & 31 & FSI (304~\AA\ and 174~\AA), METIS UV \\
           & 21:17:16–21:19:56 &  & 160 &  \\
           & 21:19:56–22:59:32 &  & 31 &  \\
\hline
2023-10-05 & 19:20:05–22:21:25 & METIS VL (tB) & $\sim$61$^{*}$ & FSI 174~\AA, METIS UV   \\
           &  &  &  & \\
\enddata
\tablecomments{\footnotesize
WB = Wideband; VL = Visible Light; UV = Ultraviolet; tB = Total Brightness; pB = Polarized Brightness. The tilde symbol ($\sim$) indicates the mode cadence (i.e., the most frequently occurring value) within the time interval. The hash symbol ($^{\#}$) denotes that at most one frame deviates from the mode, with the deviation being less than 6 seconds. The asterisk ($^{*}$) indicates multiple frames with cadence deviations from the mode, but with no more than one differing frame occurring simultaneously. The cadence variation in such cases ranges from 153 to 183 seconds.
}

\end{deluxetable*}

\section{Methodology}
\label{sec:Methodology}

\subsection{Image Preprocessing}
\label{sec:ImagePreProc}

\begin{figure*}[ht!]
    \centering
    \includegraphics[width=\textwidth]{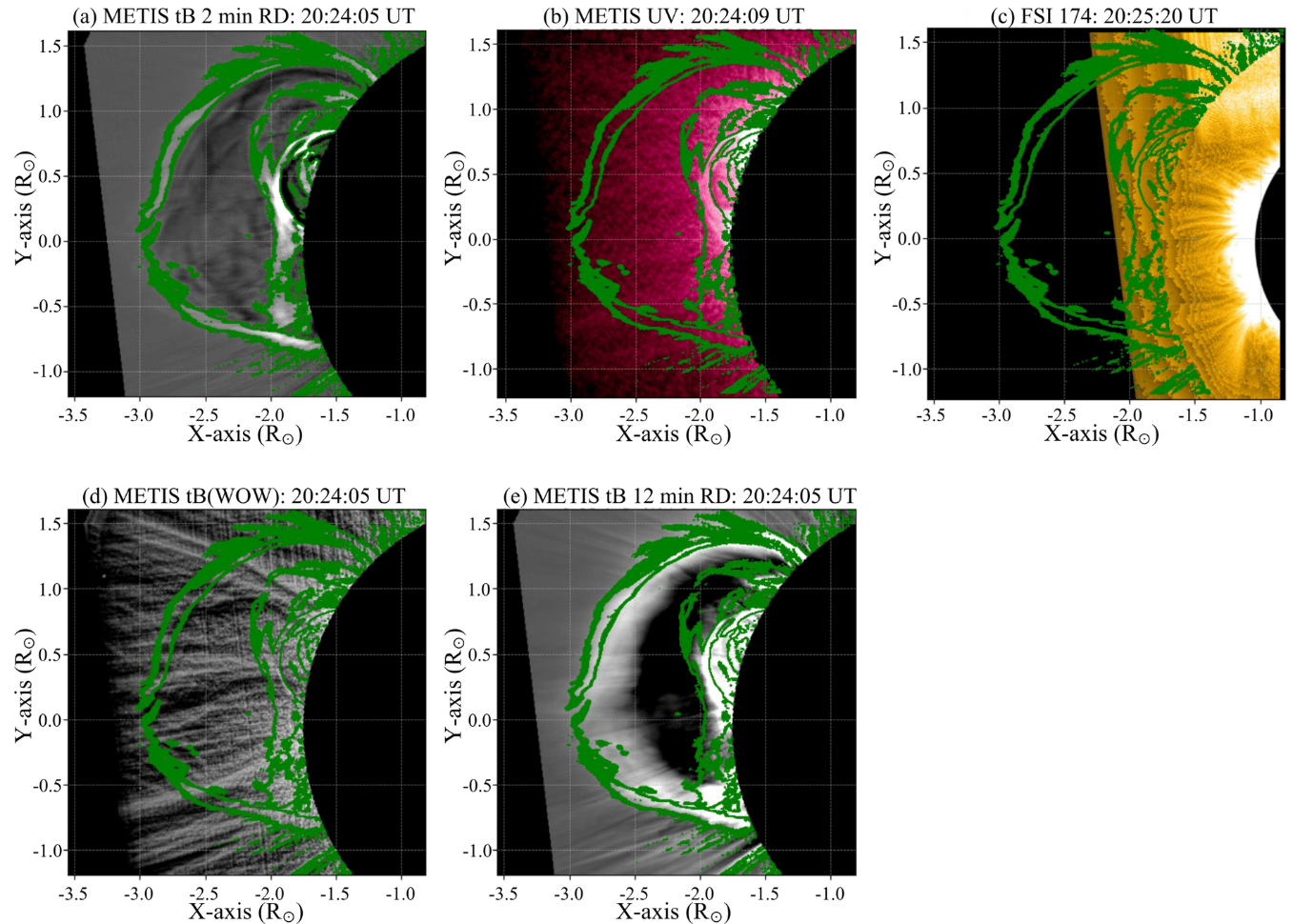}
    \caption{\footnotesize
    Demonstration of the impact of differencing interval on the intelligibility of CME features, using METIS and FSI observations from the 2023-10-05 CME event. Panel (a): METIS tB image processed with a 2-minute running difference (RD). Panel (b): Wavelets Optimized Whitening (WOW; \citealp{Auchere2023}) processed METIS UV image at a similar time, highlighting prominence material associated with the CME. Panel (c): FSI 174~\AA\ EUV image showing coronal loop structures that form the curved structure between the CME front and core. The solar disk is masked to improve contrast in the corona. Panel (d): WOW processed METIS tB frame. Panel (e): METIS tB with a 12-minute RD, where fine structures appear blurred and merged due to motion smearing. Green intensity threshold contours, derived from Panel (a), are overplotted on all other panels to demonstrate spatial correspondence of CME substructures across wavelengths and processing methods.
    }
    \label{fig:blur_demo}
\end{figure*}

To enhance the visibility of dynamic coronal transients, such as CMEs, and to suppress quasi-static background contributions, we constructed running difference (RD) sequences from total brightness (tB) and Wideband images obtained by METIS and ASPIICS, respectively. The differencing interval was fixed at two frames (i.e., $\mathrm{img}[i+2] - \mathrm{img}[i]$), which was empirically determined to optimize feature clarity. Using a shorter interval caused subtraction of the trailing edge of evolving structures, producing artificially dark artifacts. Conversely, longer intervals introduced motion blur, leading to an apparent broadening and merging of fine structures. This effect is evident in Figure~\ref{fig:blur_demo}, where a 12-minute running difference, similar to the cadence of  Large Angle Spectroscopic COronagraph C2 (LASCO C2;
\citealp{Brueckner1995}) on board the Solar and Heliospheric Observatory (SOHO; \citealp{Domingo1995}), causes internal CME features to merge and blend together, obscuring fine structural details. For the cadence available in our dataset, the two-frame interval offered the best compromise between temporal resolution and feature intelligibility. This suggests that finer substructures within CME fronts and cores can be better visualised when differencing intervals are carefully optimised to avoid artificial broadening.

Spatial masks were applied to exclude regions outside the usable field of view for each instrument, as described below. For METIS, the mask was defined by two concentric circular boundaries: an inner exclusion zone corresponding to the occulter geometry---accounting for the offset of the occulted region from the image centre due to optical misalignment, as described in the \href{https://metis.oato.inaf.it/docs/METIS-OATO-SPE-021_2.2_Solar_Orbiter_Metis_Data_Product_Description_Document.pdf}{Metis Data Product Description Document}---and an outer cut-off marking the extent of the calibrated field, both determined from known instrument parameters. In ASPIICS, where the field of view is annular due to the external occulter design, we retained regions beyond $1.6~R_\odot$ up to the detector edge, as most pixels within $1.6~R_\odot$ were saturated in the 10-second exposure channel. Additional saturated pixels beyond the $1.6~R_\odot$ radius were also masked.

\begin{figure*}[ht!]
\begin{interactive}{animation}{figure02_movie.mp4}
\centering
\gridline{
  \includegraphics[width=0.325\textwidth]{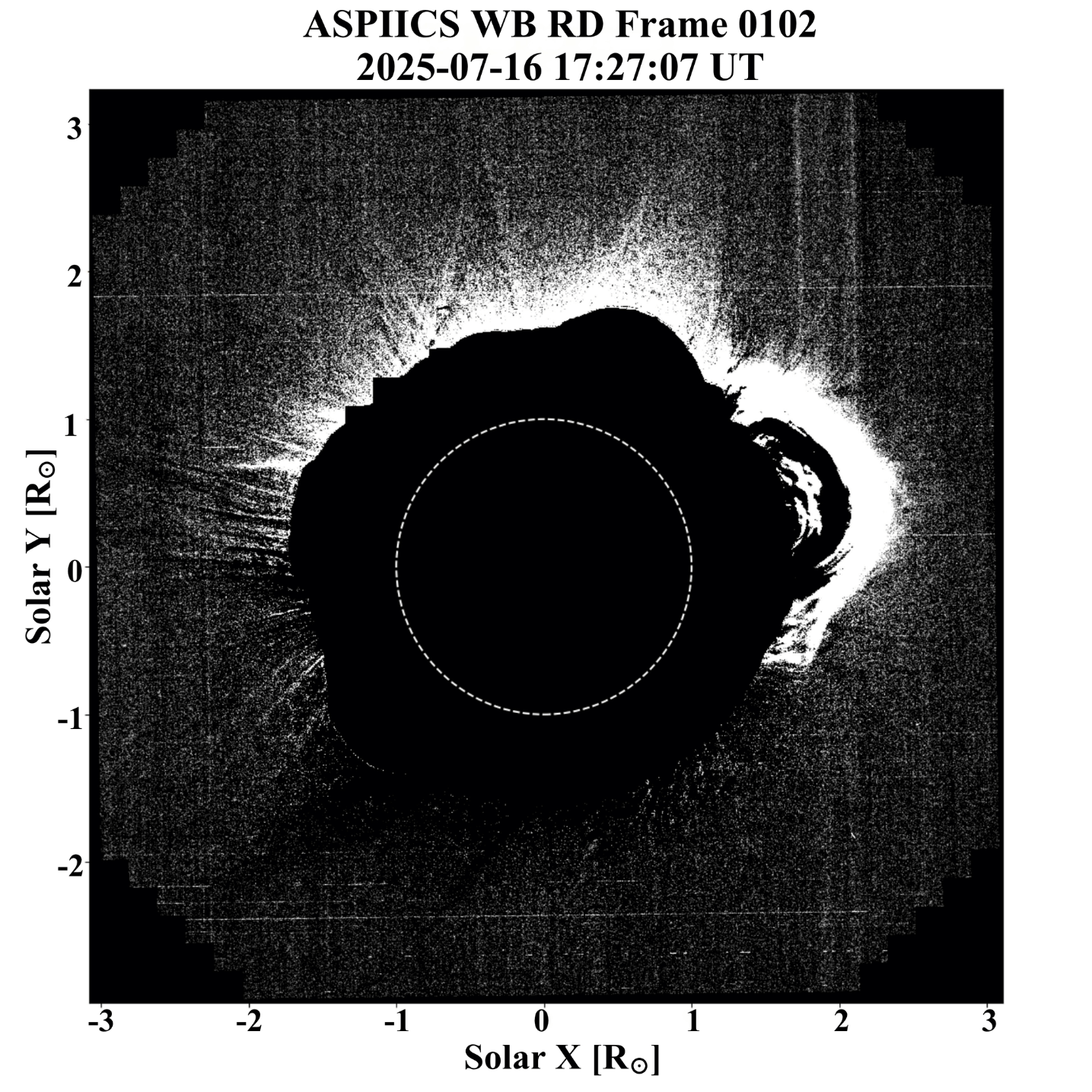}
  \includegraphics[width=0.325\textwidth]{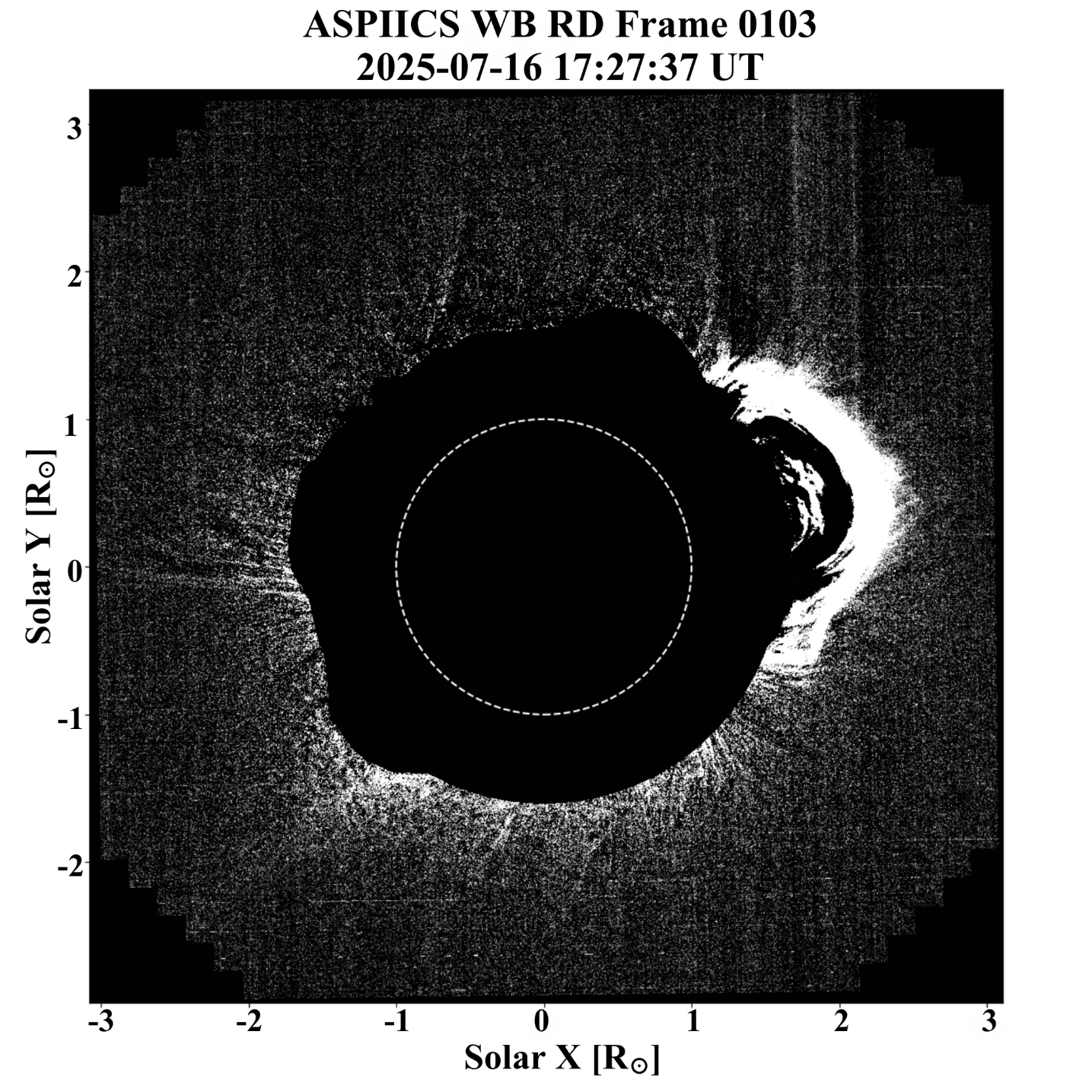}
  \includegraphics[width=0.325\textwidth]{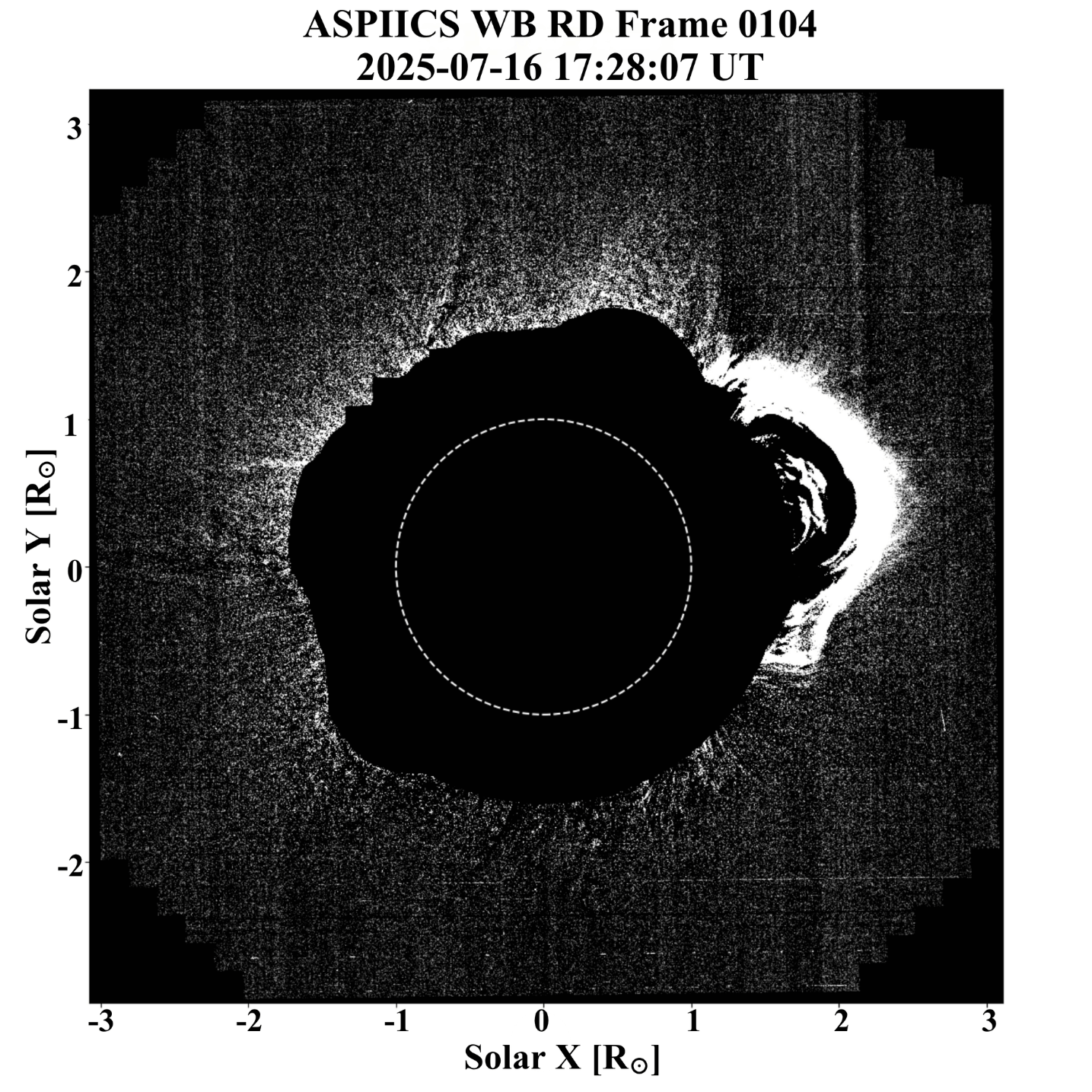}
}
\vspace{-5mm}
\gridline{
  \includegraphics[width=0.325\textwidth]{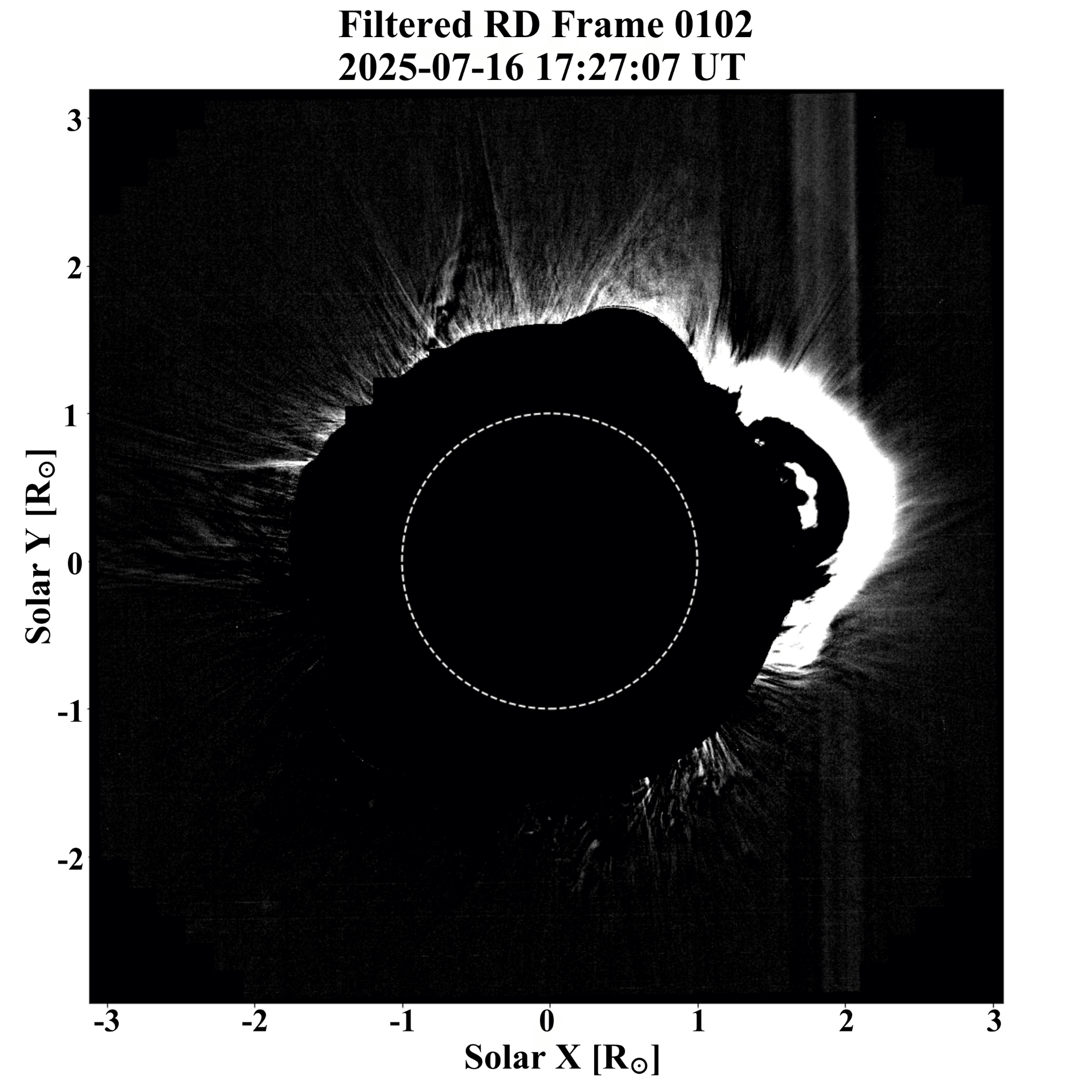}
  \includegraphics[width=0.325\textwidth]{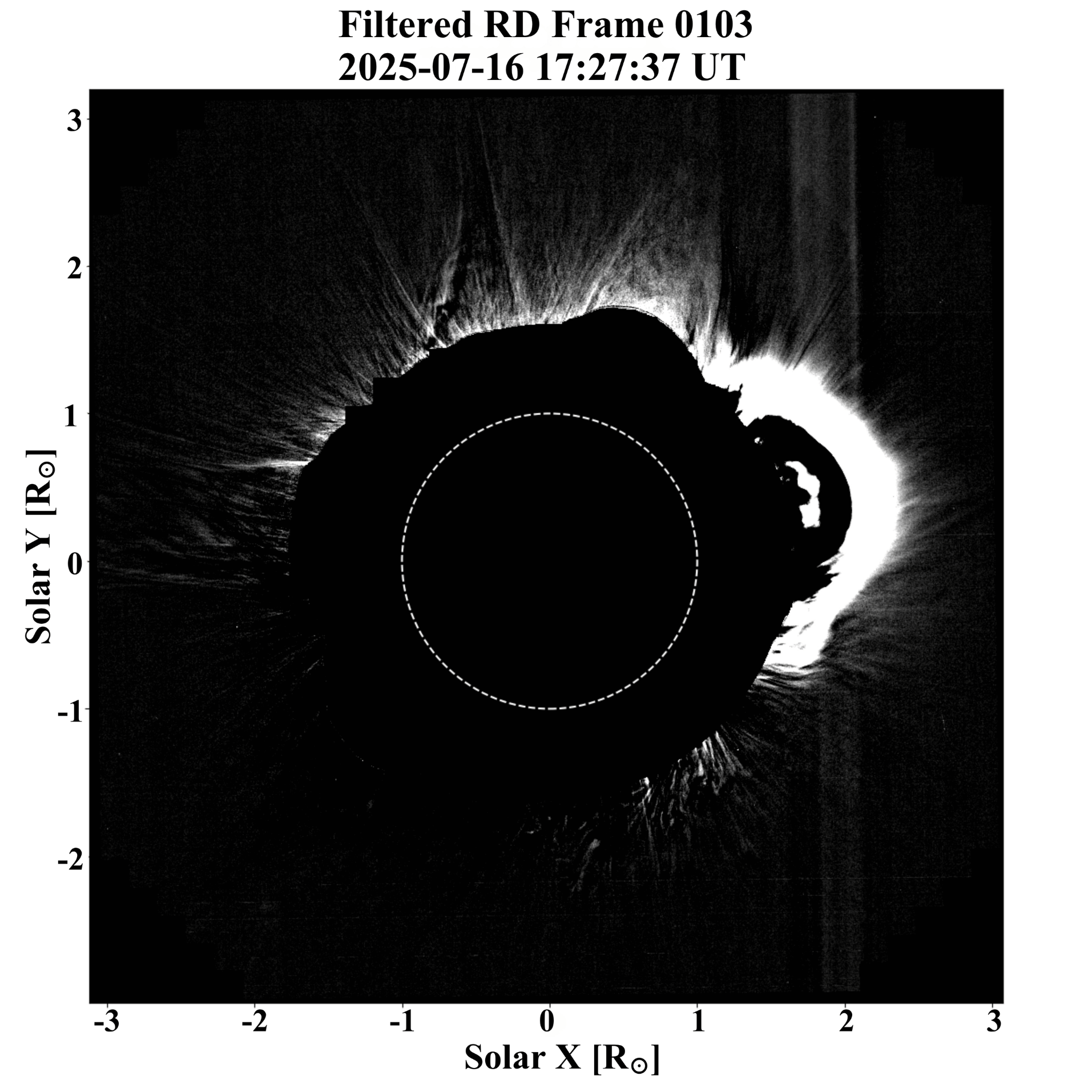}
  \includegraphics[width=0.325\textwidth]{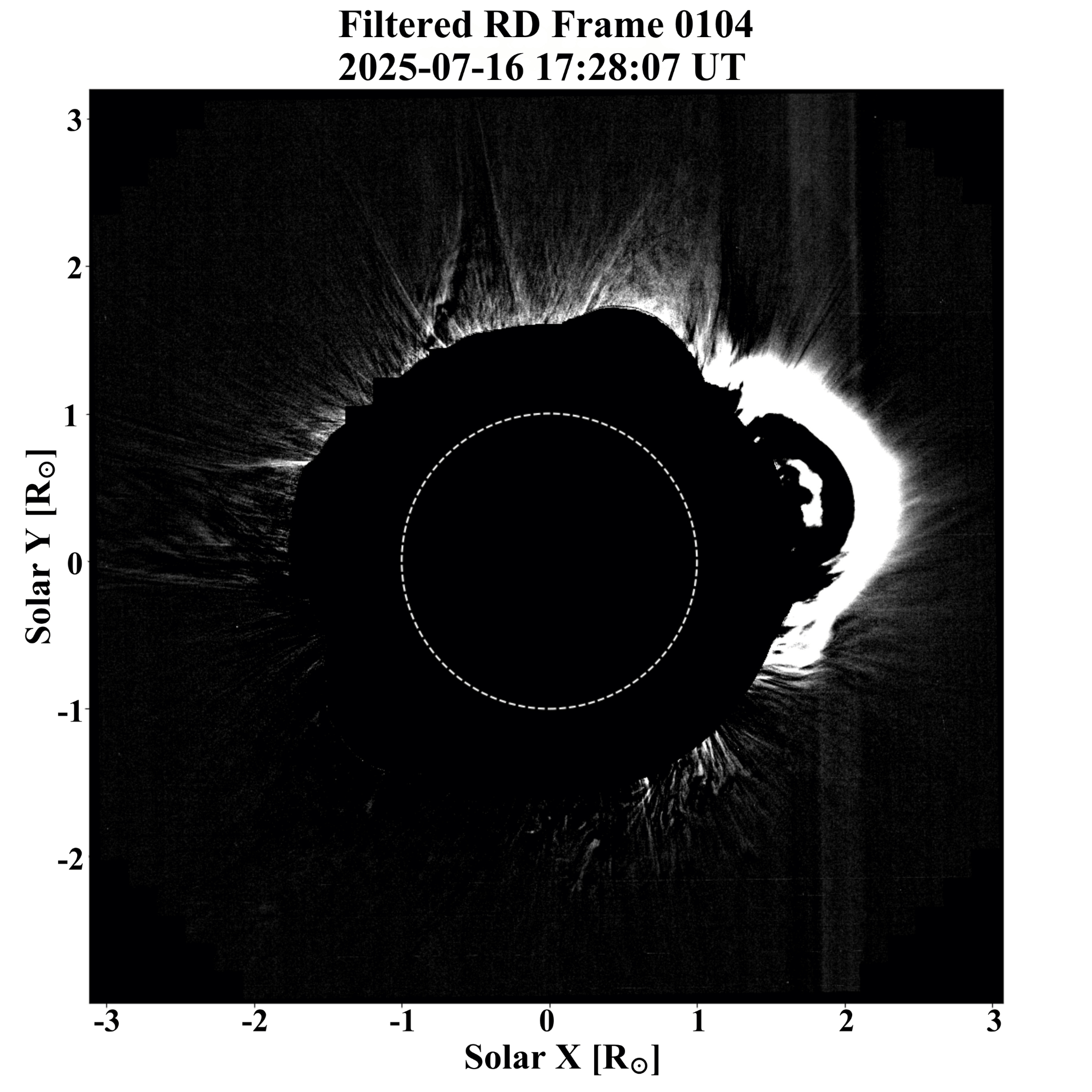}
}
\end{interactive}
\caption{\footnotesize
Demonstration of temporal flicker suppression in ASPIICS running difference images. The X and Y axes of each image are in helioprojective solar radii ($R_\odot$) units. Top row: unfiltered running difference sequence frames showing frame-to-frame brightness flickering, especially around the mask. Bottom row: corresponding frames after GTF filtering. The filtered sequence shows reduced brightness flickering and noise. Each column shows matched timestamps across the two rows for direct comparison. This figure is available as an animation. The animation starts on 2025 July 16 at 16:54 and ends at 17:34 with a duration of 4 s.
}
\label{fig:aspiics_flicker_removal}
\end{figure*}

ASPIICS data exhibited frame-to-frame flickering in brightness, as discussed in Section~\ref{sec:data_source}. This artifact, though sub-pixel in nature, introduced artificial intensity fluctuations around the occulter, which propagated into the running-difference sequences. Consequently, the optical flow algorithm often misinterpreted these fluctuations as apparent motion. To suppress this, we formulated a Gaussian-tapered Fourier (GTF) filter and applied it along the temporal axis of the image cube for each pixel. For every pixel, the intensity time series was transformed into the frequency domain using a Fourier transform. In this representation, rapid frame-to-frame brightness fluctuations appear as high-frequency components, while relatively slowly evolving structures, such as CMEs, are represented by low-frequency components. A Gaussian weighting function was then applied in the frequency domain, centred at zero frequency, such that higher frequencies were progressively down-weighted rather than abruptly removed. This “Gaussian taper” ensures a smooth transition between retained and suppressed frequencies. The filtered signal was then transformed back to the time domain to reconstruct the image sequence. A hard-cut high-pass filter was avoided, as it introduces ringing artifacts near CME locations due to abrupt truncation of high-frequency components. In contrast, the Gaussian taper provides smooth attenuation, minimising temporal artifacts while retaining coherent motion. The cut-off frequency was set to 15\% of the Nyquist limit ($f_\mathrm{cut}/f_\mathrm{Nyq} = 0.15$), with the Gaussian standard deviation set equal to the cut-off ($\sigma = f_\mathrm{cut}$). The effect of this filtering, before and after application, is illustrated in Figure~\ref{fig:aspiics_flicker_removal}, which shows the improvement in background stability and the enhanced clarity of the CME structure.

After differencing and filtering, the ASPIICS frames were corrected for roll angle. All difference images were then converted to helioprojective coordinates and scaled in solar radii ($R_\odot$) for physical consistency.

To suppress background fluctuations and enhance the contrast of faint CME features, we applied an exponential intensity normalization across the image sequence. This involved computing the global intensity range and scaling each image by fixed exponential factors derived from the global minimum and maximum. The transformation preserved the original pixel distribution shape but non-linearly compressed lower-intensity values—typically associated with background noise—while maintaining the relative contrast of brighter CME structures.

These preprocessed and normalized sequences served as the input for the optical flow and kinematic analyses detailed in the following sections.

\subsection{Optical Flow Estimation}
\label{sec:optical_flow}

To estimate the plane-of-sky (POS) motion of coronal mass ejection (CME) substructures across successive frames, we employed a dense optical flow algorithm based on the Farnebäck method \citep{Farneback2003}. Optical flow techniques infer pixel-wise displacement vectors between consecutive images under the assumption of brightness constancy, enabling the retrieval of detailed two-dimensional velocity fields.

Before computing optical flow, the running difference image sequences (Section~\ref{sec:Methodology}) were denoised using the bilateral filter \citep{Tomasi1998}, which combines spatial and intensity-domain weighting to suppress small-scale noise while preserving sharp intensity gradients. Unlike Gaussian smoothing, which blurs both noise and structural edges, bilateral filtering preserves the fine-scale boundaries of CME features, which are critical for accurate motion estimation.

For each RD image pair, the optical flow field was computed between frame~$i$ and frame~$i{+}1$ using the \texttt{calcOpticalFlowFarneback} function from the python's OpenCV library. The resulting horizontal and vertical components ($u$, $v$) were derived, which describe pixel-wise displacements. The velocity magnitude in pixel units was calculated as $V_{\mathrm{pix}} = \sqrt{u^2 + v^2}$, and converted to physical units using:
\begin{equation}
V_{\mathrm{km/s}} = \frac{V_{\mathrm{pix}} \cdot s}{\Delta t}
\end{equation}
where $s$ is the physical scale per pixel (in $\mathrm{km\ pixel^{-1}}$), derived from the instrument plate scale and spacecraft–Sun distance, and $\Delta t$ is the inter-frame time in seconds.

To reduce the influence of noise and artefacts, we applied lower and upper thresholds to the optical flow velocities. A lower threshold of 50~$\mathrm{km\ s^{-1}}$ was used to avoid detecting background noise as flow. An upper threshold of 3000~$\mathrm{km\ s^{-1}}$ was imposed to exclude unrealistically large velocities arising from transient image artefacts, such as glare or sudden large-scale intensity changes between frames. These thresholds were chosen based on empirical tests and ensured that such artefacts did not bias the averaged velocity estimates. In particular, omitting the lower threshold resulted in average velocities that were $\sim$10-16 times lower than those obtained after applying the threshold.

\begin{figure*}[ht!]
\centering
\gridline{
  \fig{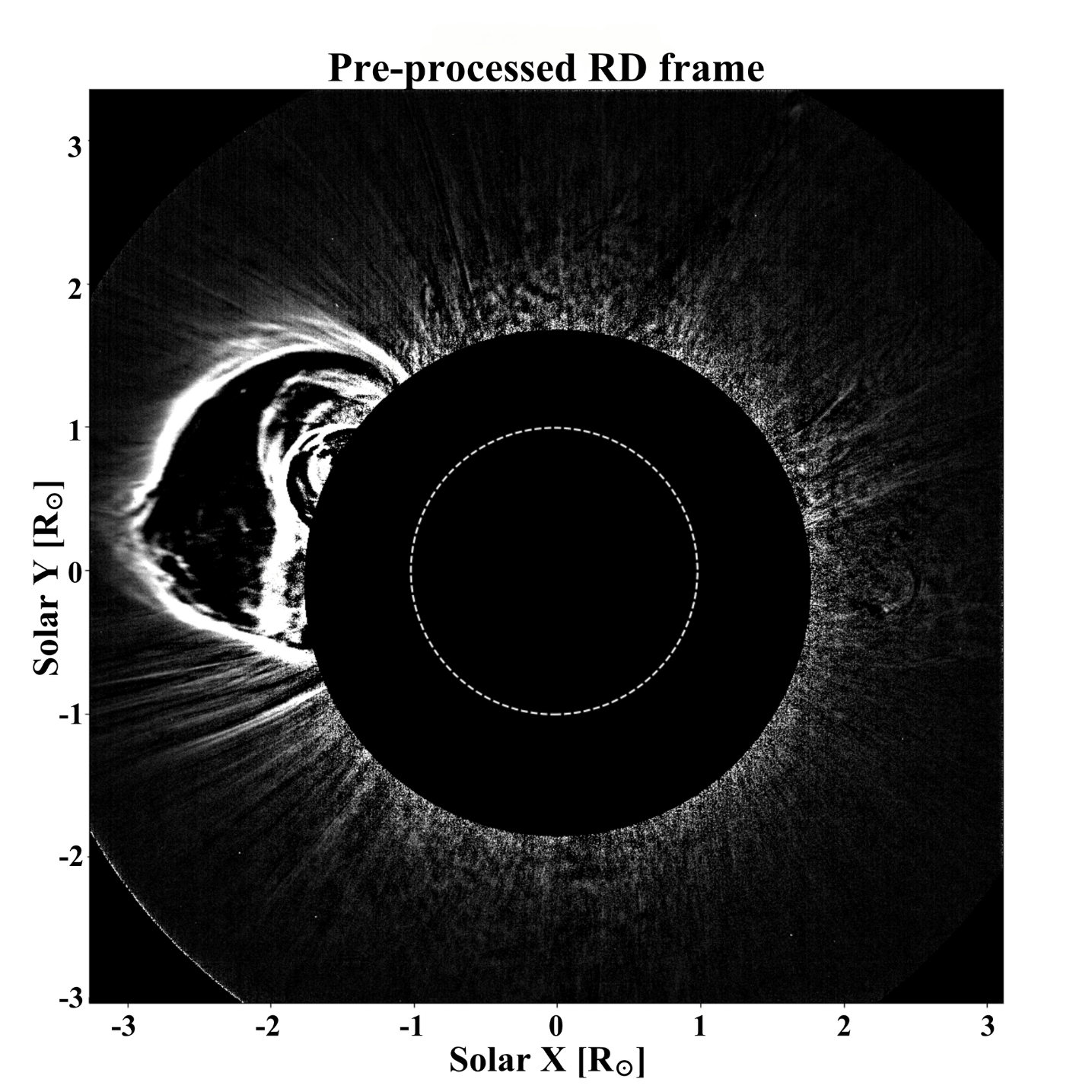}{0.48\textwidth}{(a) Preprocessed running difference}
  \fig{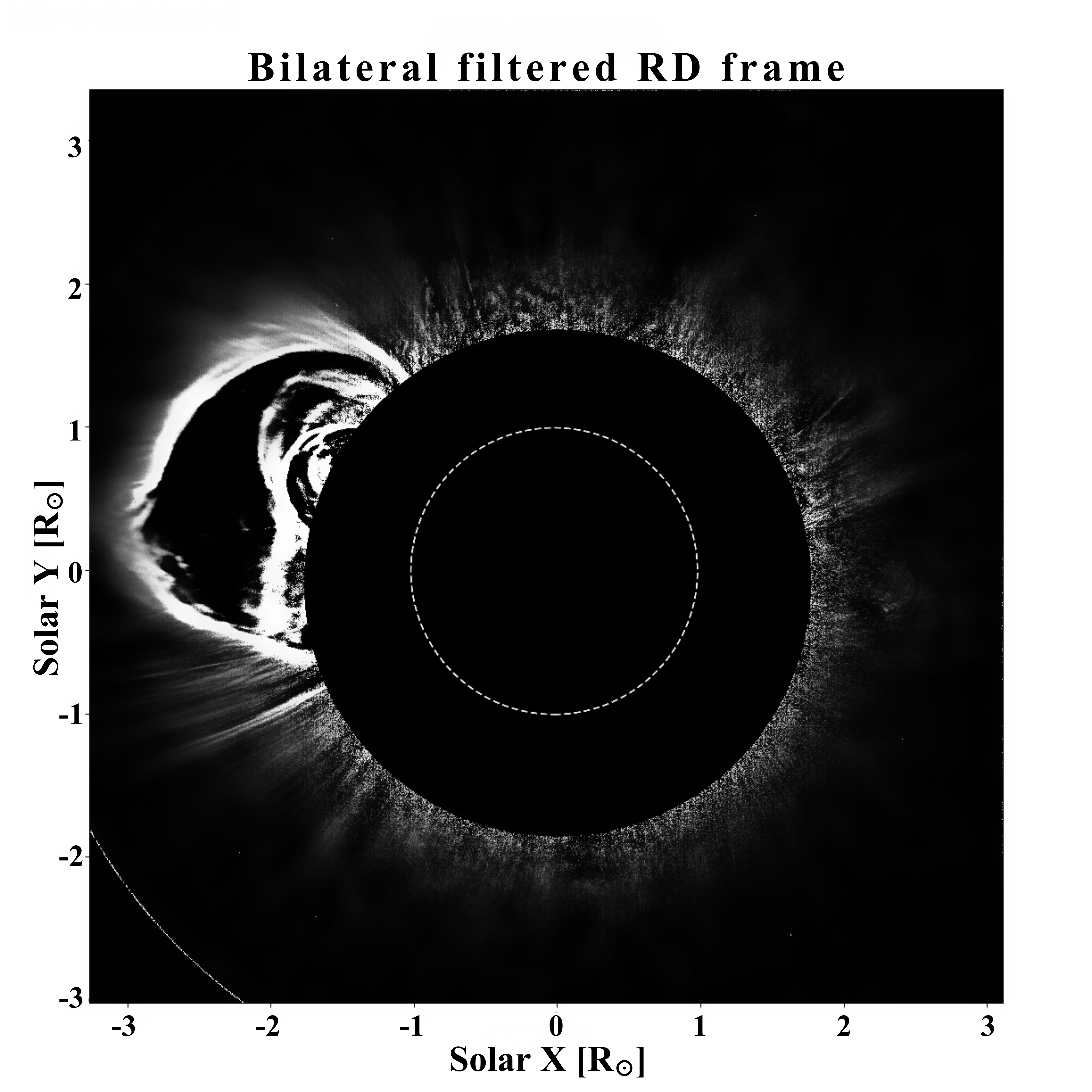}{0.48\textwidth}{(b) Bilateral filtered image}
}
\gridline{
  \fig{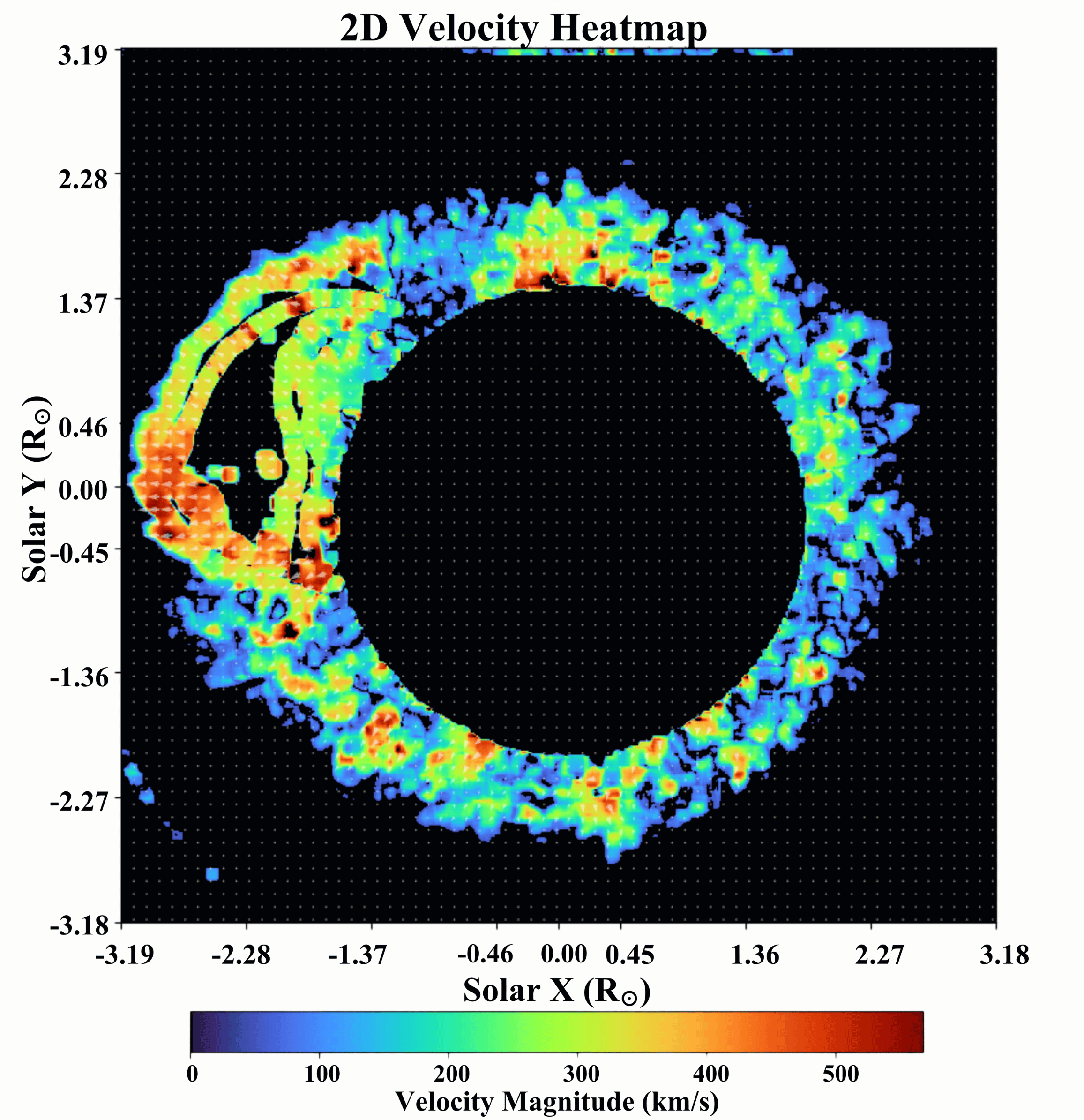}{0.485\textwidth}{(c) Velocity magnitude heatmap}
  \fig{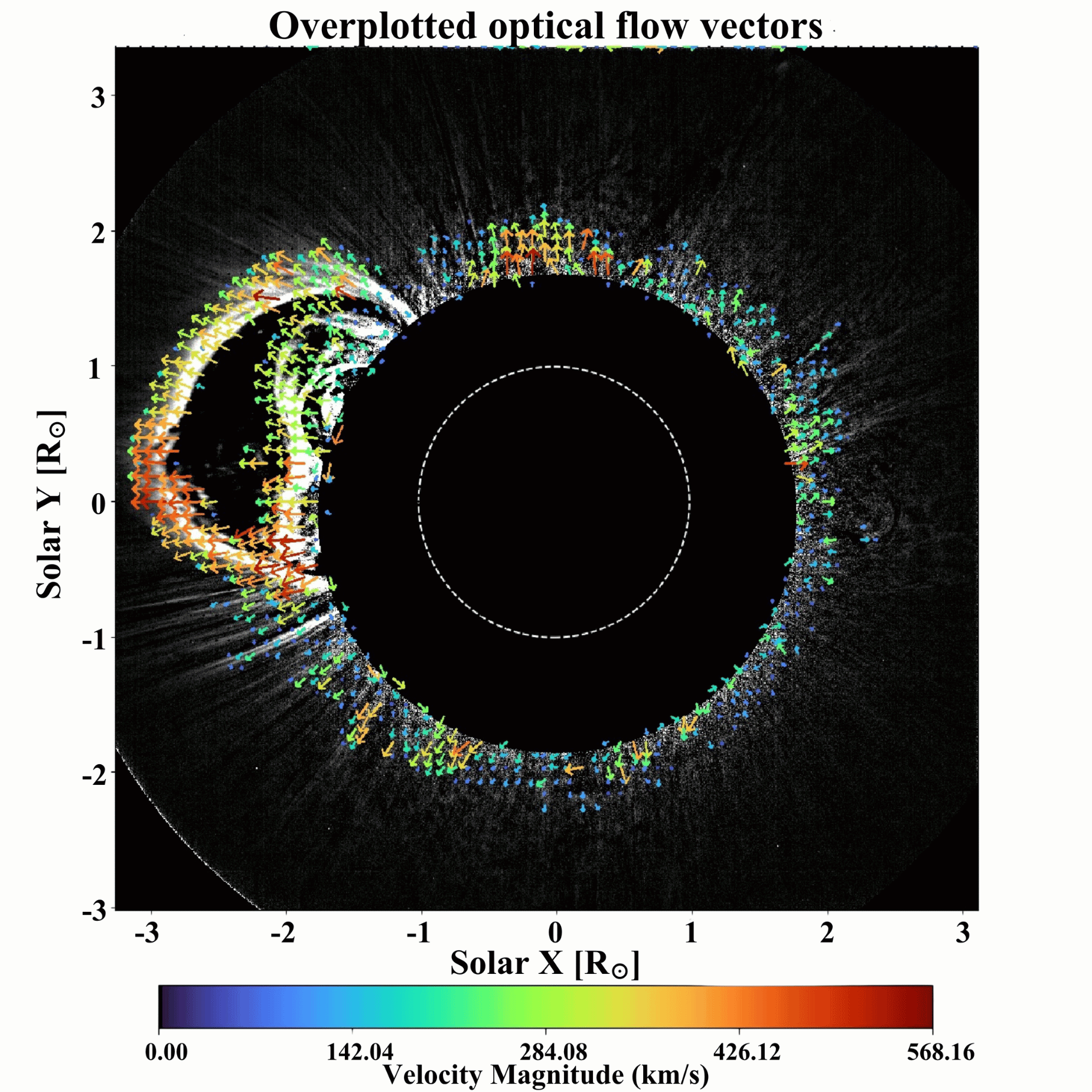}{0.49\textwidth}{(d) Velocity vectors on RD image}
}
\caption{\footnotesize
visualisation of the pipeline for estimating CME flow fields from METIS running difference images: (a) preprocessed running-difference frame after masking, (b) result after bilateral filtering to reduce noise while preserving CME edges, (c) velocity magnitude heatmap computed using dense optical flow, and (d) velocity vectors overplotted on the original frame, color-coded by speed. The arrows indicate the spatial location of features in the subsequent frame. All panels are shown in helioprojective coordinates and scaled in solar radii ($R_\odot$).
}
\label{fig:flow_viz_pipeline}
\end{figure*}

The velocity fields were visualised as direction-encoded arrows and velocity magnitude heatmaps, both scaled in solar radii using helioprojective coordinates. Figure~\ref{fig:flow_viz_pipeline} illustrates the complete processing pipeline, which forms the basis for the kinematic analysis in Section~\ref{sec:Results}. The methodology described in this section constitutes DOFCAT (Dense Optical Flow CME Analysis Tool).

\section{Results}
\label{sec:Results}

\subsection{Validation of the Optical Flow pipeline}
\label{sec:flow_validation}

To validate the optical flow-derived velocity fields, we compared the CME front speed at a specific position angle (PA) with an independent measurement based on a time–distance (XT) map.

\begin{figure*}[ht!]
\centering

\gridline{
  \fig{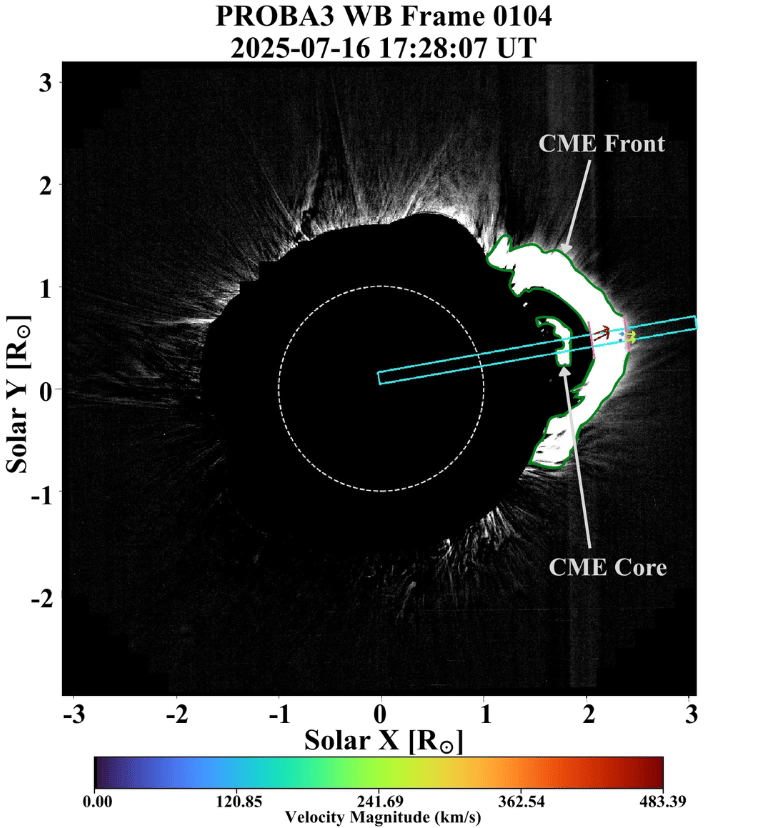}{0.32\textwidth}{(a)}
  \fig{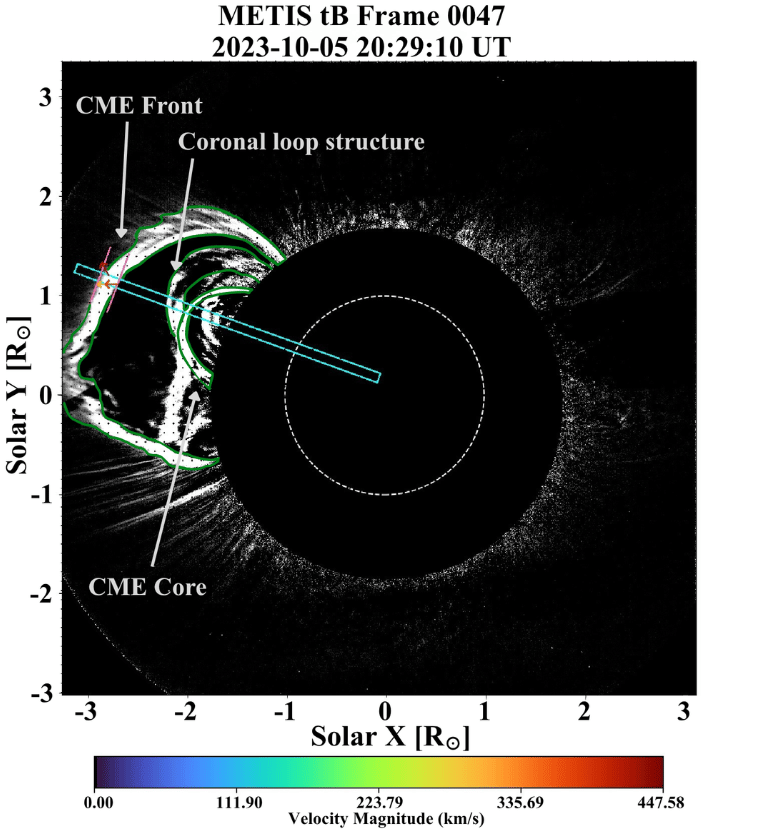}{0.32\textwidth}{(b)}
  \fig{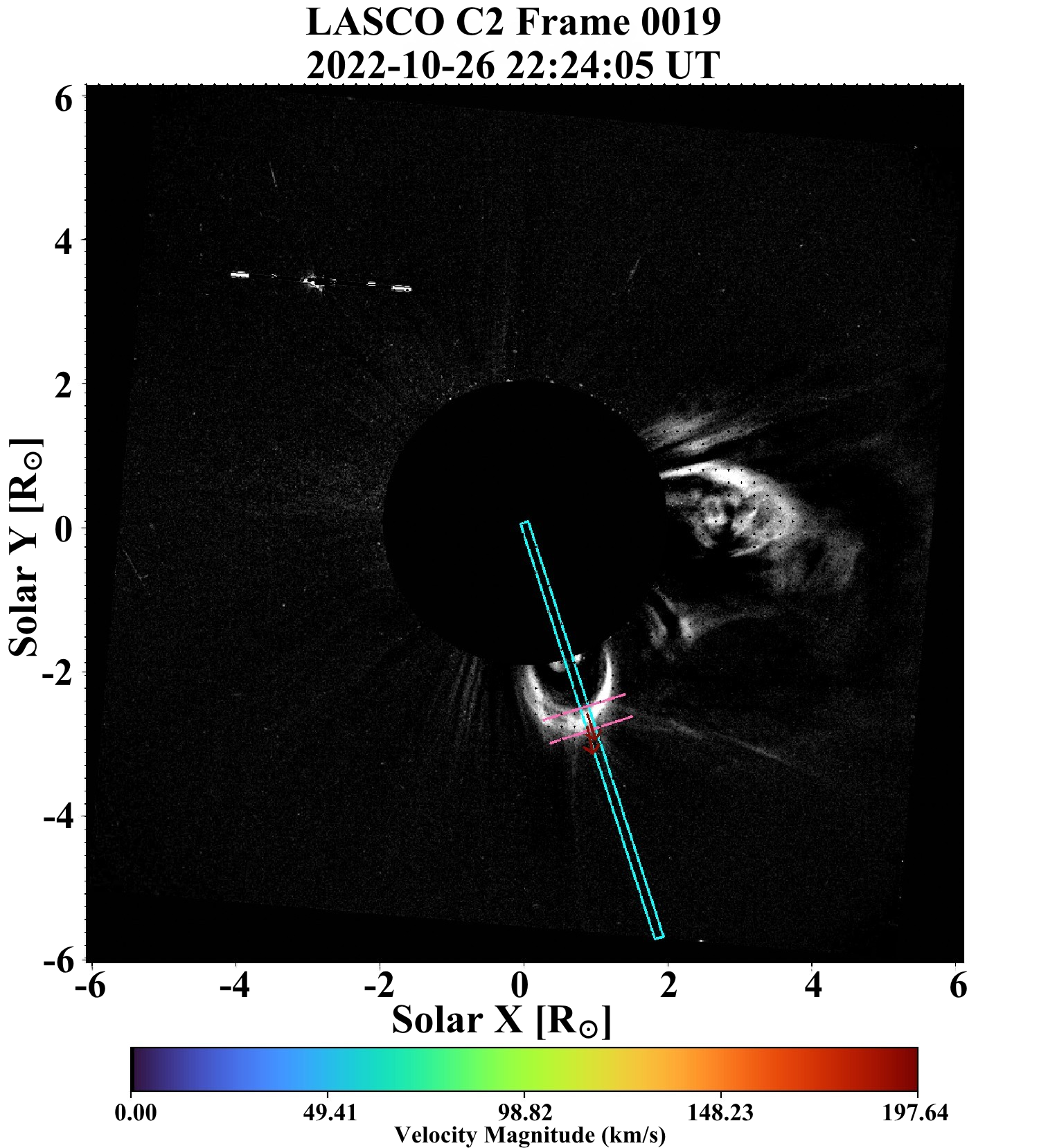}{0.32\textwidth}{(c)}
}

\vspace{-3mm}

\gridline{
  \fig{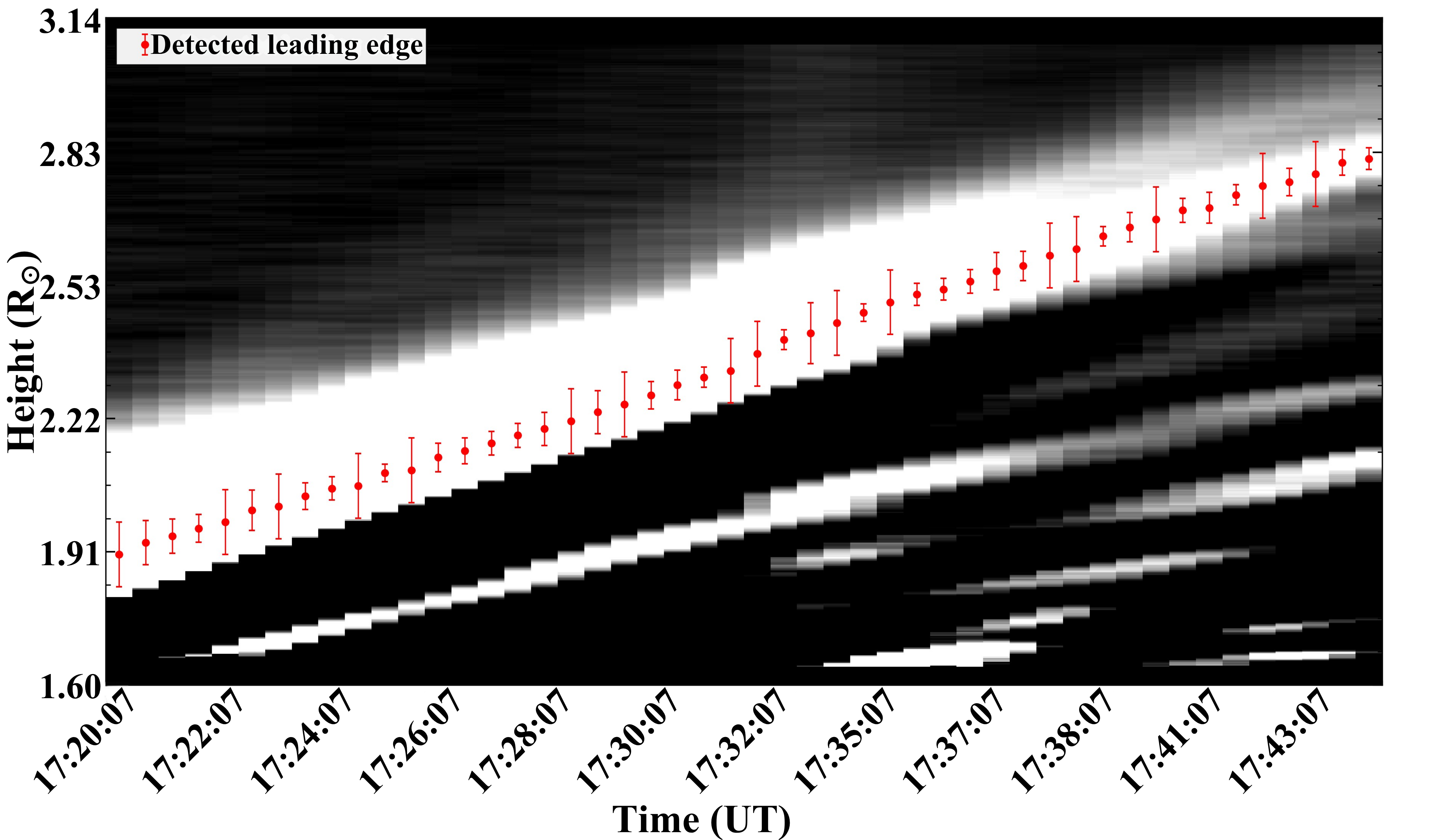}{0.325\textwidth}{(d)}
  \fig{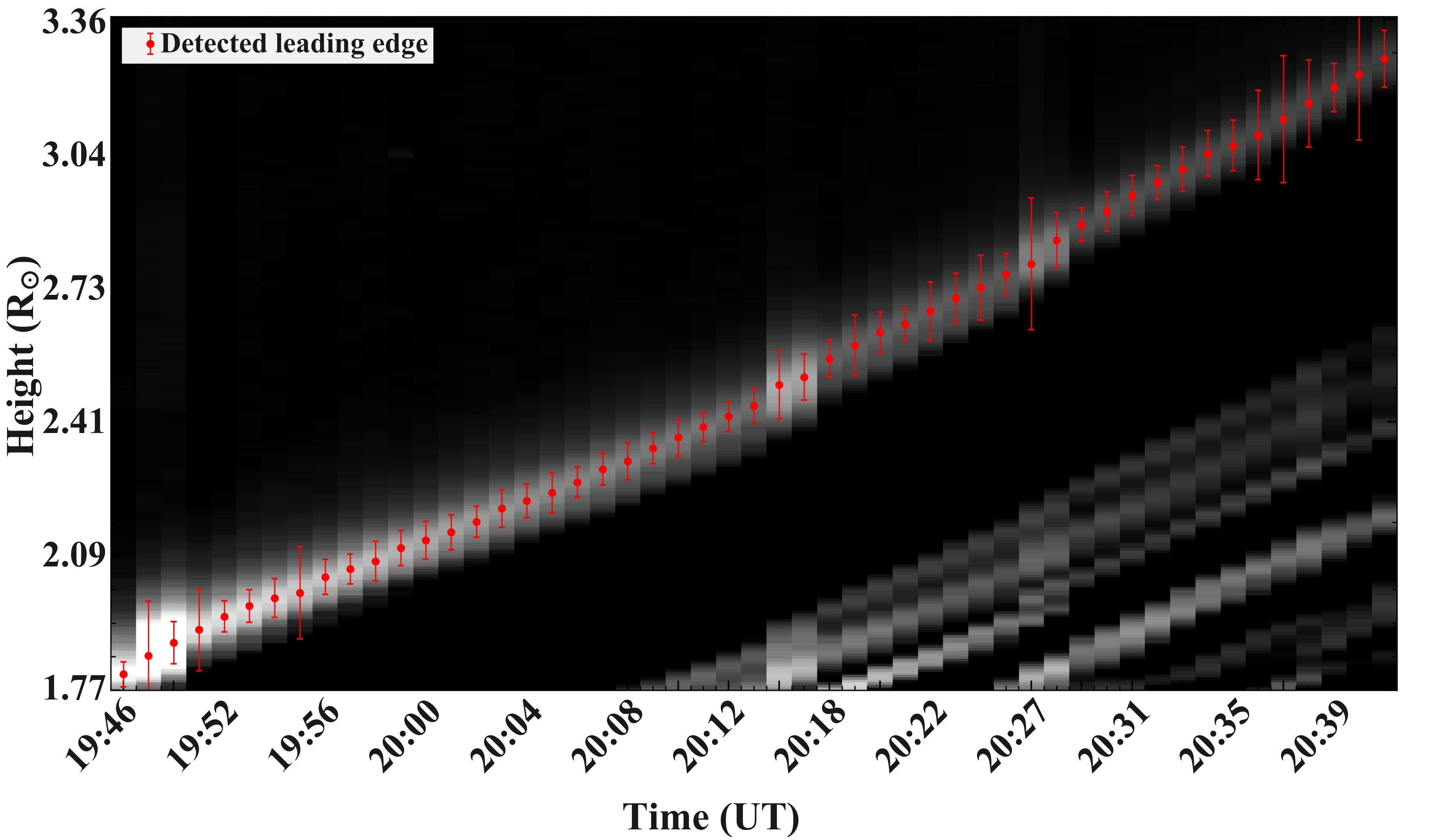}{0.325\textwidth}{(e)}
  \fig{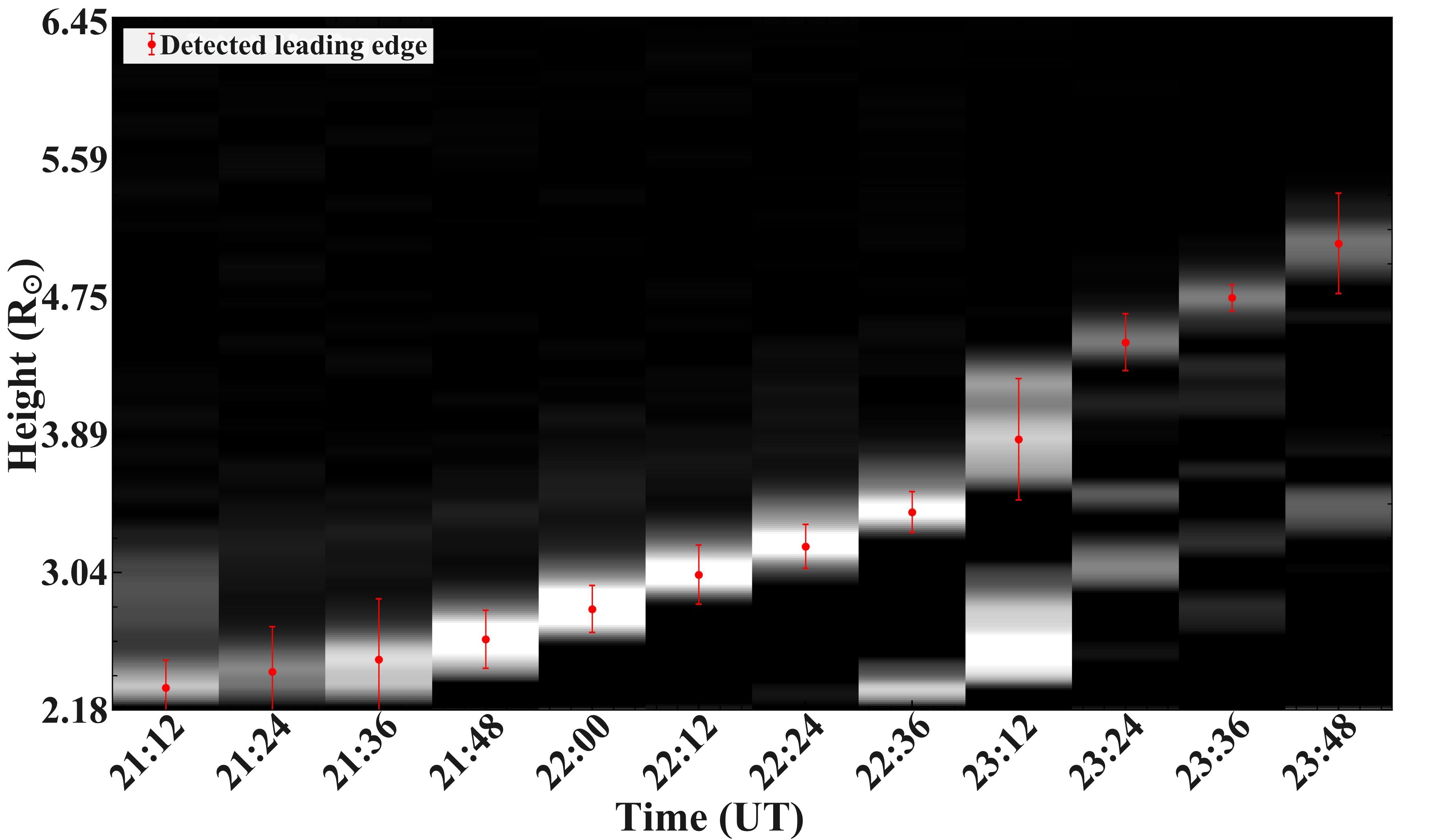}{0.325\textwidth}{(f)}
}

\vspace{-5mm}

\gridline{
  \fig{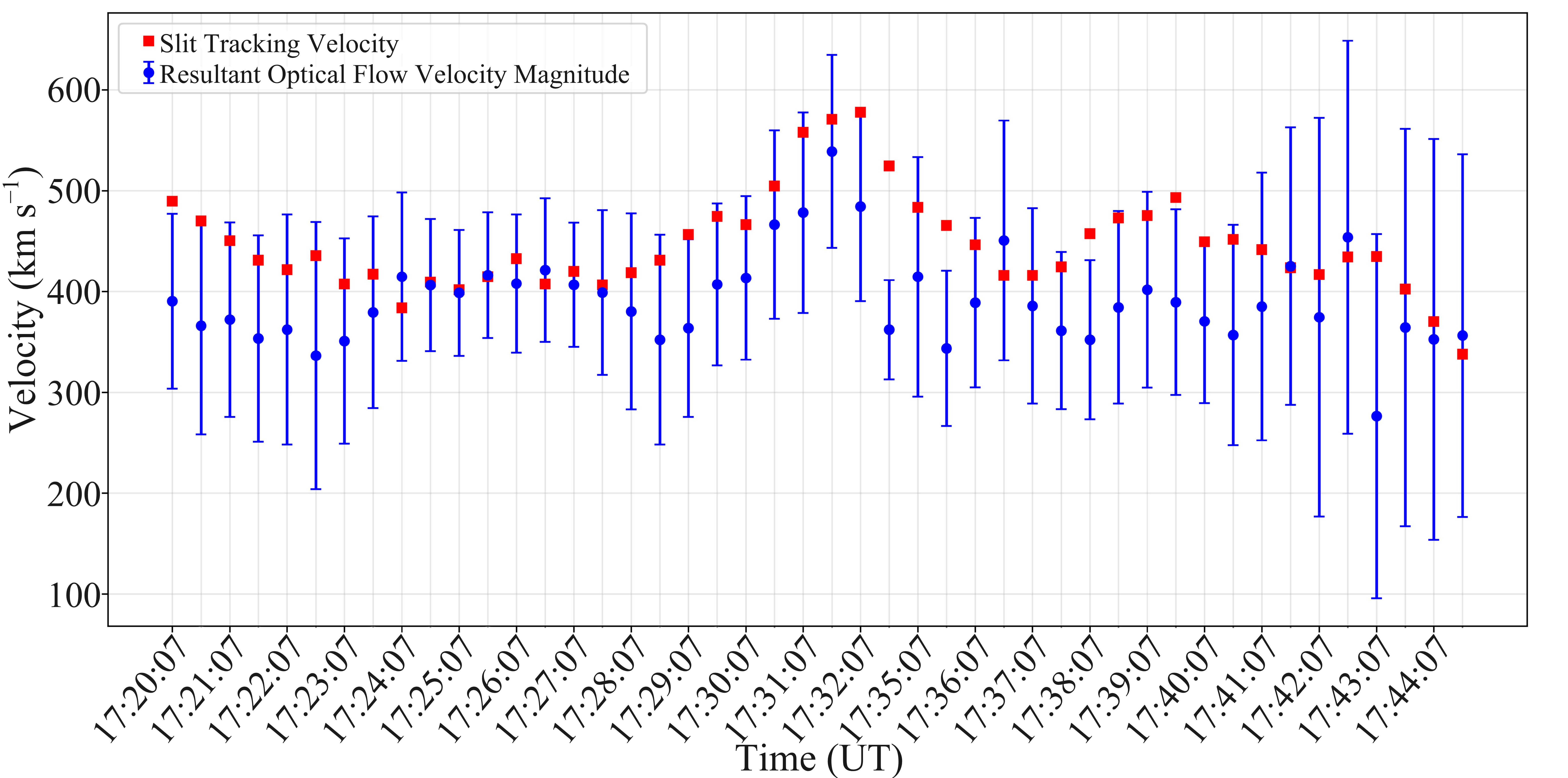}{0.325\textwidth}{(g)}
  \fig{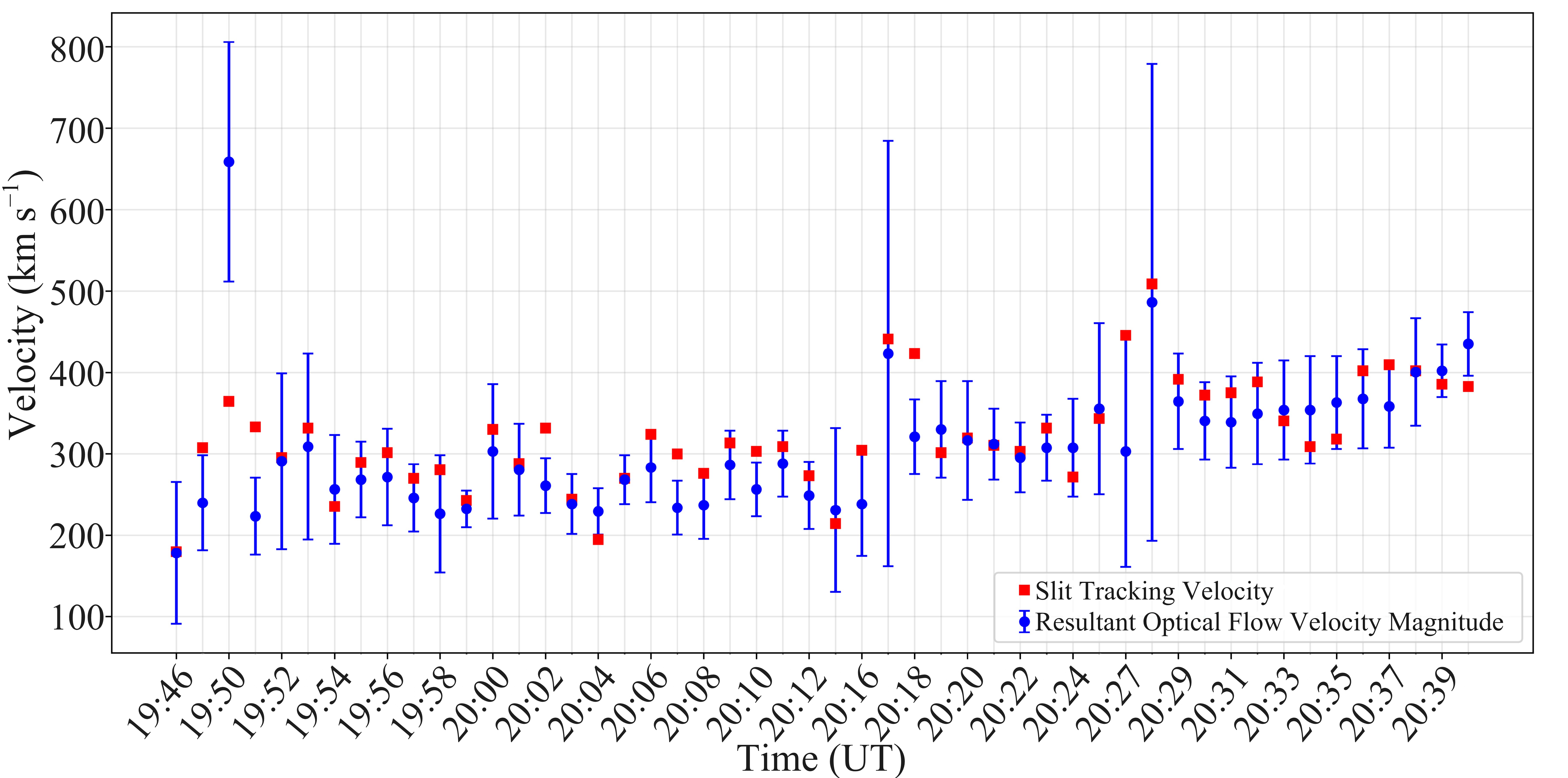}{0.325\textwidth}{(h)}
  \fig{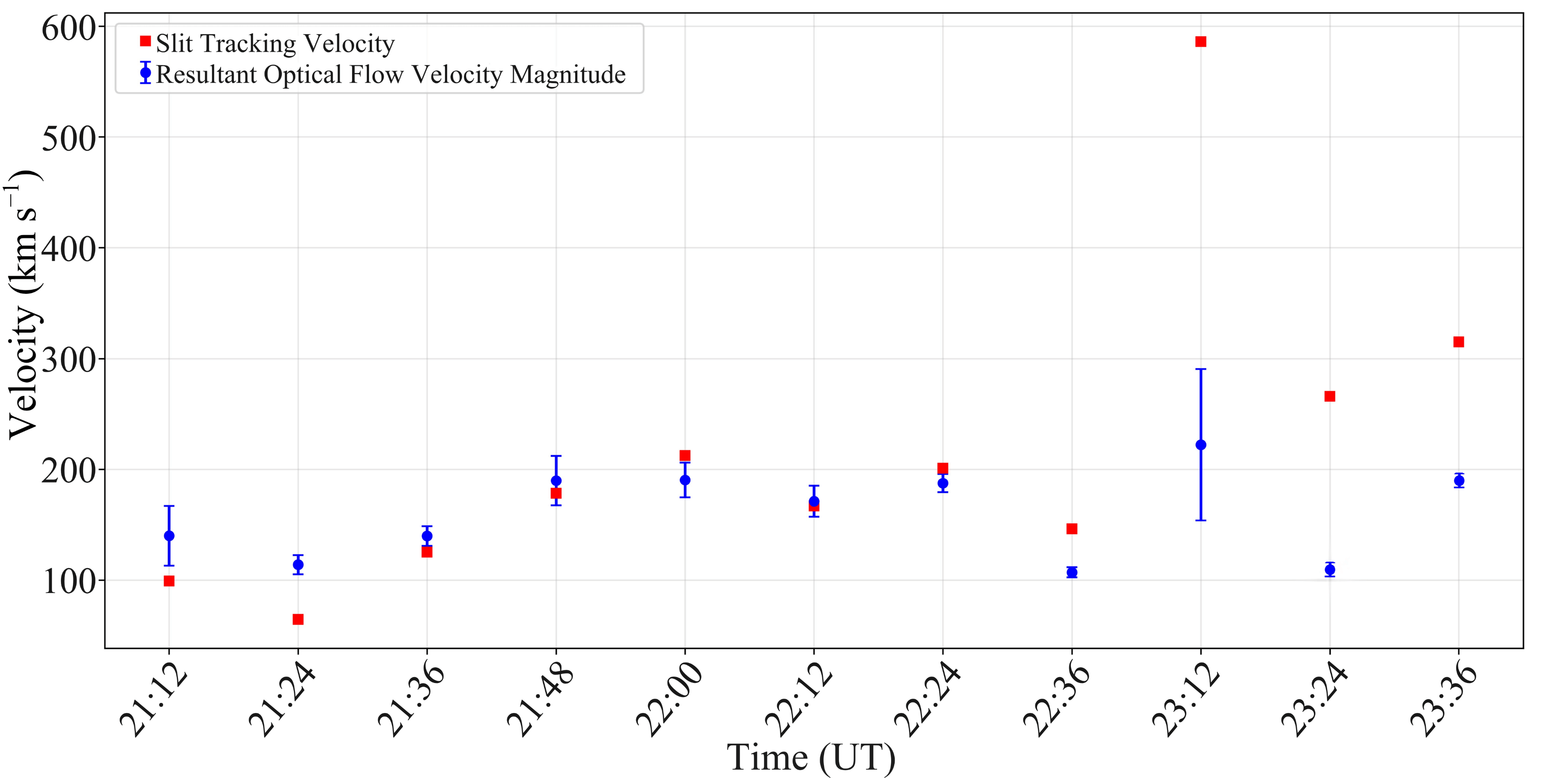}{0.325\textwidth}{(i)}
}

\caption{\footnotesize
Validation of optical flow–derived velocities using slit-based tracking across ASPIICS (2025-07-16) in the first column, METIS (2023-10-05) in the second column, and LASCO C2 (2022-10-26) in the third column. 
Top row (a–c): selected running difference frames from each instrument with a fixed radial slit overlaid along 278°, 70°, and 195° position angles (PA), respectively, intersecting the CME front. The pink-highlighted region along the slit denotes the CME front analysis zone. The CME substructures (front, core, and coronal loop where present) are manually delineated using green contours in panels (a) and (b). Middle row (d–f): corresponding time–distance (XT) maps generated along the slit for each instrument, with the detected CME front traced with red points. The error bars denote the standard deviation of the Gaussian fitted intensity profiles. 
Bottom row (g–i): velocity–time profiles comparing slit-based tracking (red) with optical flow–derived magnitudes (blue), with error bars representing the standard deviation of optical flow vectors within the selected region.
}
\label{fig:xt_vs_flow_validation}
\end{figure*}

To construct the XT map, we placed a radial slit of width between 15 to 20 pixels, extending along a chosen PA as shown in the top row of Figure~\ref{fig:xt_vs_flow_validation}. The slit was averaged across its width to enhance the signal-to-noise ratio, and the resulting intensity profiles were stacked over time to generate the time–distance map. The CME front was tracked using a semi-automated peak-intensity detection algorithm applied along each time slice of the XT map (middle row of Figure~\ref{fig:xt_vs_flow_validation}). The process was initiated manually by selecting the time-step in which the front first became clearly visible. For each subsequent frame, the algorithm identified the local intensity maximum along the radial height column of the XT map, corresponding to the brightest feature of the CME front. To ensure that the detected height reflected the outward motion of the CME front, a constraint was imposed such that the detected position at each time step was required to lie above that of the previous time step. This reduced false detections from downflows or background features. The extracted height–time profile was subsequently smoothed using a Savitzky–Golay filter \citep{Savitzky1964} with a sliding window size of 5 and polynomial order of 2. \citet{Byrne2013} and \citet{Vashishtha2023} demonstrated the effectiveness of using the Savitzky-Golay filter to reduce high-frequency noise while preserving the underlying CME kinematics. The velocity was subsequently calculated as $\Delta h / \Delta t$. This method was selected to maintain consistency with the optical flow approach, which also derives motion based on intensity displacement between adjacent frames.

For comparison, optical flow velocity vectors were sampled along the same radial slit and over the same time interval, and the resulting flow velocities were extracted. Both methods showed good agreement in magnitude and exhibited similar temporal trends as evident in the bottom row of Figure~\ref{fig:xt_vs_flow_validation}, supporting the reliability of the optical flow technique for retrieving CME kinematics from running difference coronagraph sequences. To further verify the robustness of the approach, we applied the same method to LASCO C2 observations as illustrated in the third column of Figure~\ref{fig:xt_vs_flow_validation}.

On average, the magnitudes of the resultant velocities derived from optical flow were lower than the corresponding slit-based velocities by approximately 11–13\% for ASPIICS and METIS, and by about 30\% for LASCO C2. This systematic underestimation likely reflects a difference between the two methods. The slit-based technique tracks the position of the brightness point along the CME front at a given position angle, whereas the optical-flow velocity is computed as the spatial mean of the flow field over an extended region of the front at the same position This region contains a spread of velocities, quantified by the standard deviation shown as error bars in our velocity measurements, and averaging over this spread naturally pulls the mean below the single-point value obtained from the slit method.

\subsection{Velocity Dispersion Between CME Substructures}
\label{sec:velocity_dispersion_between}

To investigate the kinematic evolution of CME substructures, we applied localized optical flow analysis along narrow radial slits at selected position angles (PAs) and time frames where distinct features were visually distinguishable and simultaneously present. The selection of the PAs were prioritised around the apex of the CME front. In each frame, three regions were manually marked along the slit: Region~1 (leading front), Region~2 (core), and Region~3 (coronal loop, if present). Optical flow vectors within each region were extracted, where each vector comprises horizontal ($u$) and vertical ($v$) components (Section~\ref{sec:optical_flow}). The $u$ and $v$ components were averaged independently over all pixels in the region, and the resultant velocity magnitude was then computed from these averaged components. The plane-of-sky (POS) projected height of each region was similarly tracked by measuring the radial distance to its frontmost edge in every frame. While our primary objective was to examine the kinematic coupling between substructures, these height profiles also enabled a comparison with the results reported by \citet{Majumdar2024}, where they studied the evolution of LE–core height separation in the inner corona.

The resulting height and velocity–time profiles were smoothed using a Savitzky-Golay filter (with the same parameters as in Section~\ref{sec:flow_validation}) to suppress noise while preserving the temporal behaviour of each region. To quantify internal structural separation, we computed the relative height offsets of the core and loop's leading edge with respect to the CME front's trailing edge.

\begin{figure*}[ht!]
\centering
\gridline{
  \fig{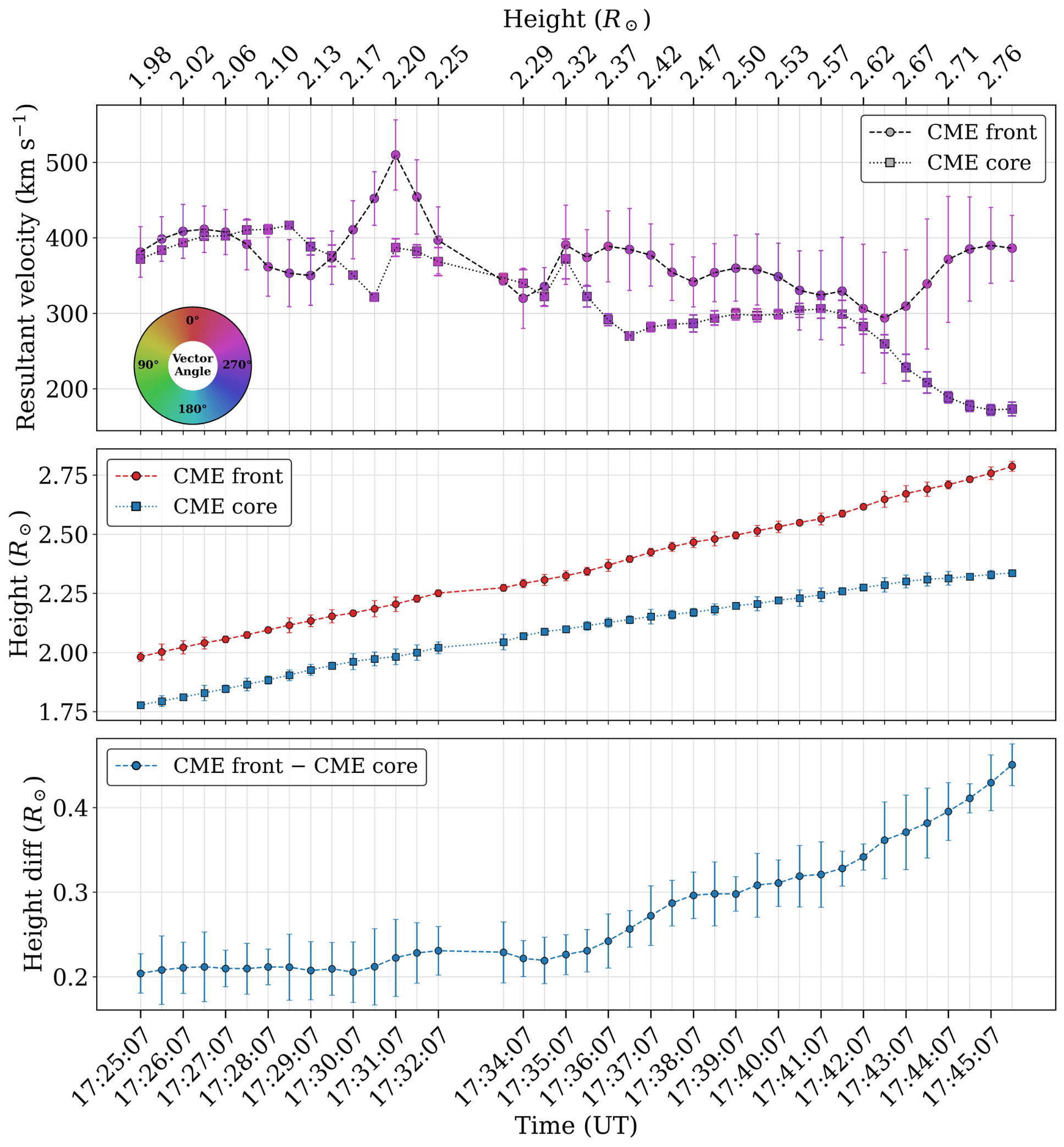}{0.48\textwidth}{(a) PA = 280°}
  \fig{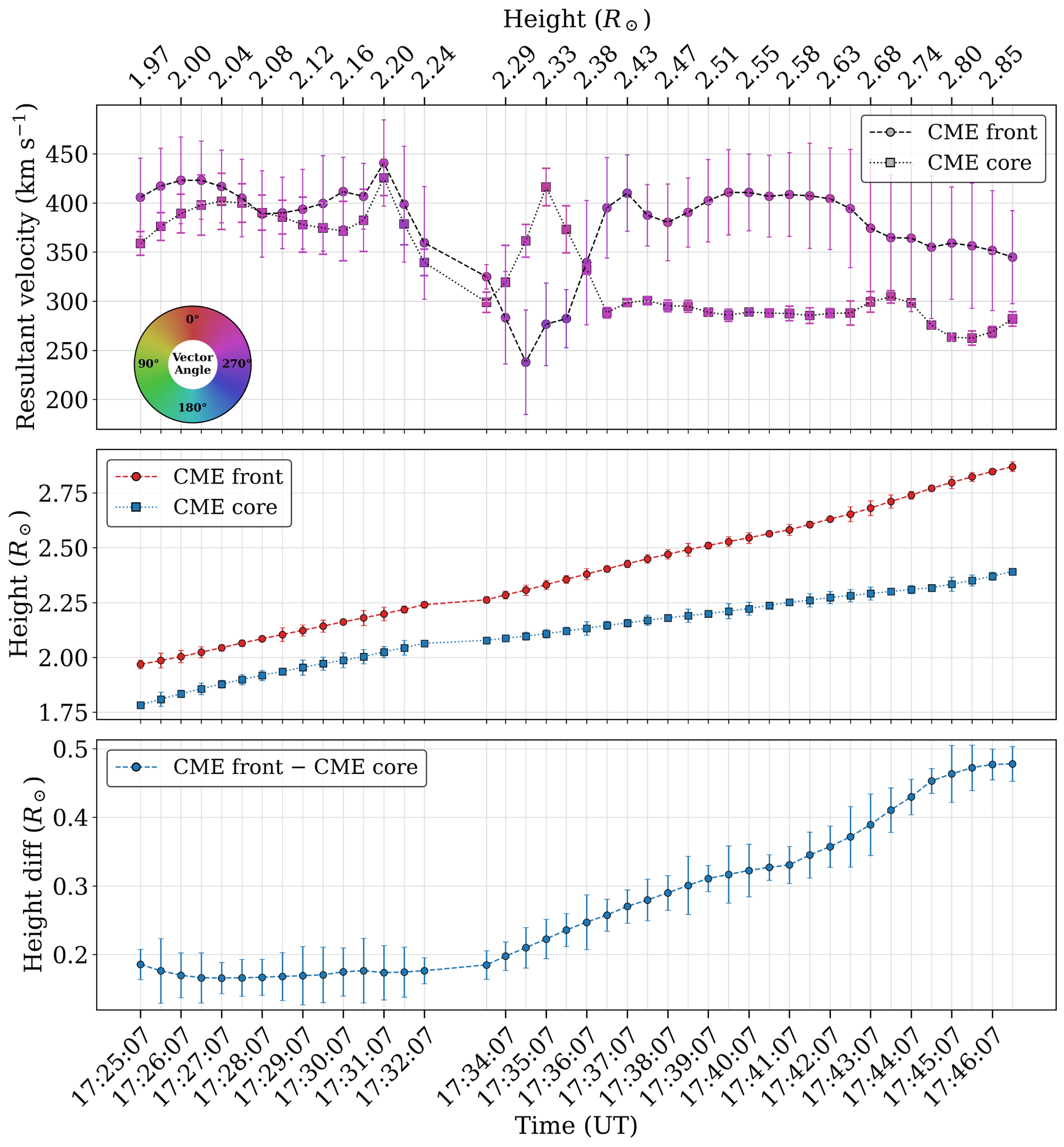}{0.48\textwidth}{(b) PA = 283°}
}

\vspace{-2mm}

\centerline{
  \includegraphics[width=0.48\textwidth]{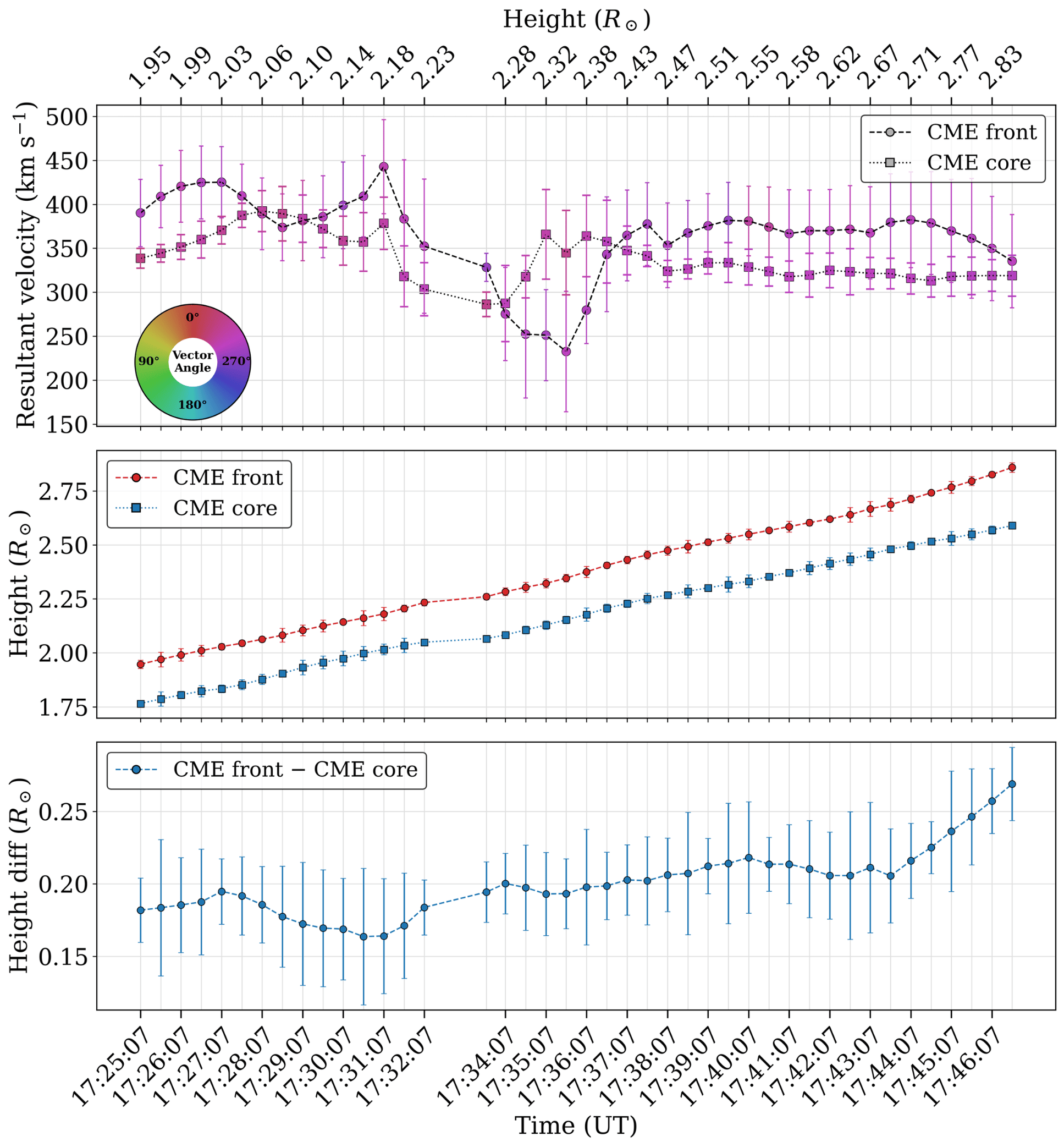}
}
\vspace{-1mm}
\centerline{(c) PA = 285°}

\vspace{-2mm}
\caption{\footnotesize
Substructure kinematic diagnostics for the 2025 July 16 CME observed by ASPIICS at three position angles (PAs). Each panel shows three stacked plots: (top) time evolution of the resultant flow velocity for the CME front and core; (middle) radial height of each substructure in units of $R_\odot$. The error bars are derived from the standard deviation of ten repeated manual point selections; (bottom) height difference between the front and core over time. In the top panels, marker colours indicate the direction of the resultant optical flow velocity vectors, as encoded by the hue wheel inset, where each colour corresponds to a specific propagation angle in the plane of the sky. The dashed and dotted lines connecting data points are included solely for visual clarity and do not represent fitted curves. The upper x-axis in the velocity plots provides the corresponding helioprojective height of the CME front.
}

\label{fig:aspiics20250716}
\end{figure*}
\begin{figure*}[ht!]
\centering

\gridline{
\hspace{0.18\textwidth}
\fig{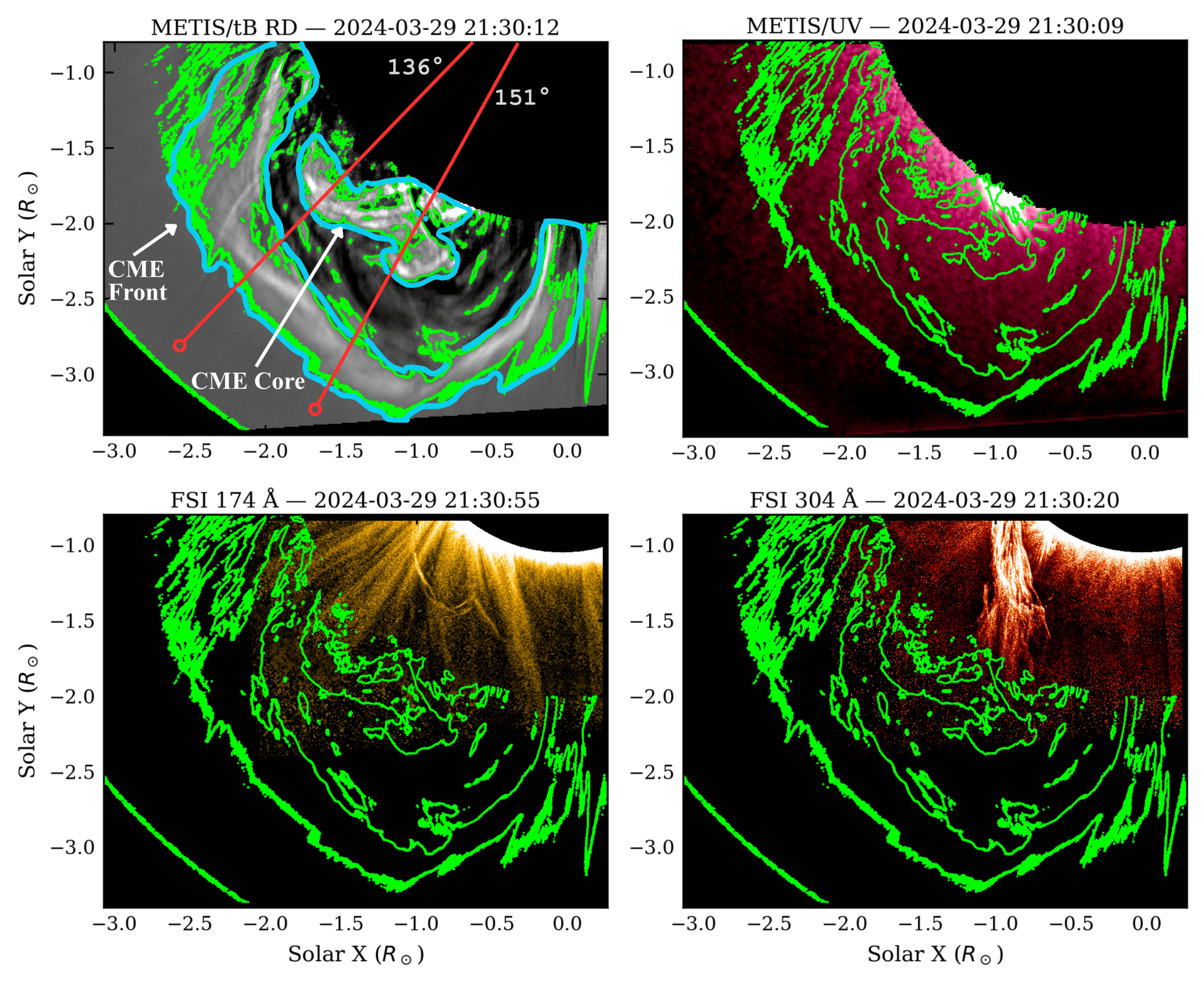}{0.55\textwidth}{(a) METIS tB RD intensity contours overplotted on METIS UV, FSI 174 \AA\ and 304 \AA\ channels}
\hspace{0.18\textwidth}
}

\vspace{2mm}

\gridline{
\fig{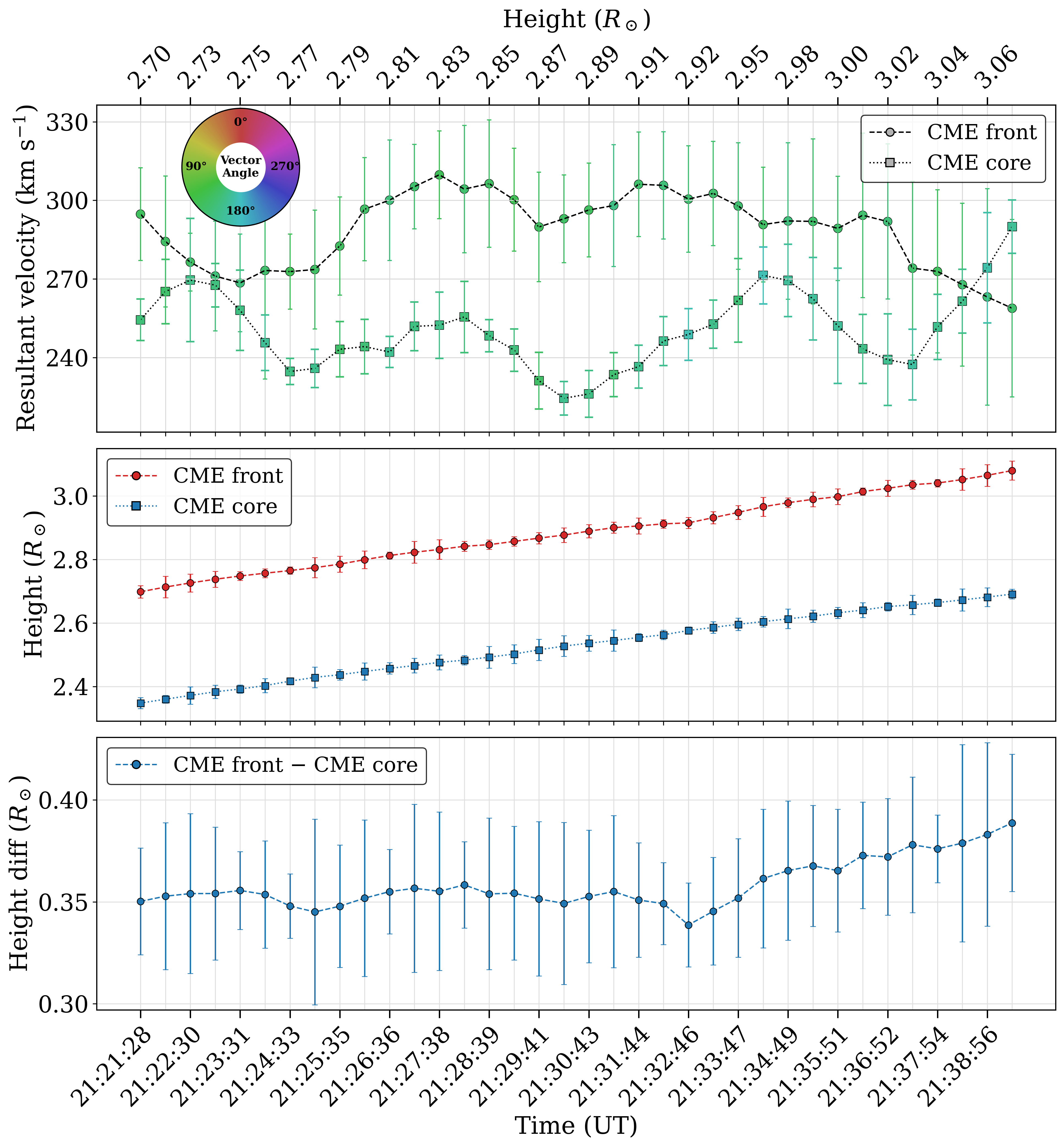}{0.48\textwidth}{(b) PA = 136°}
\fig{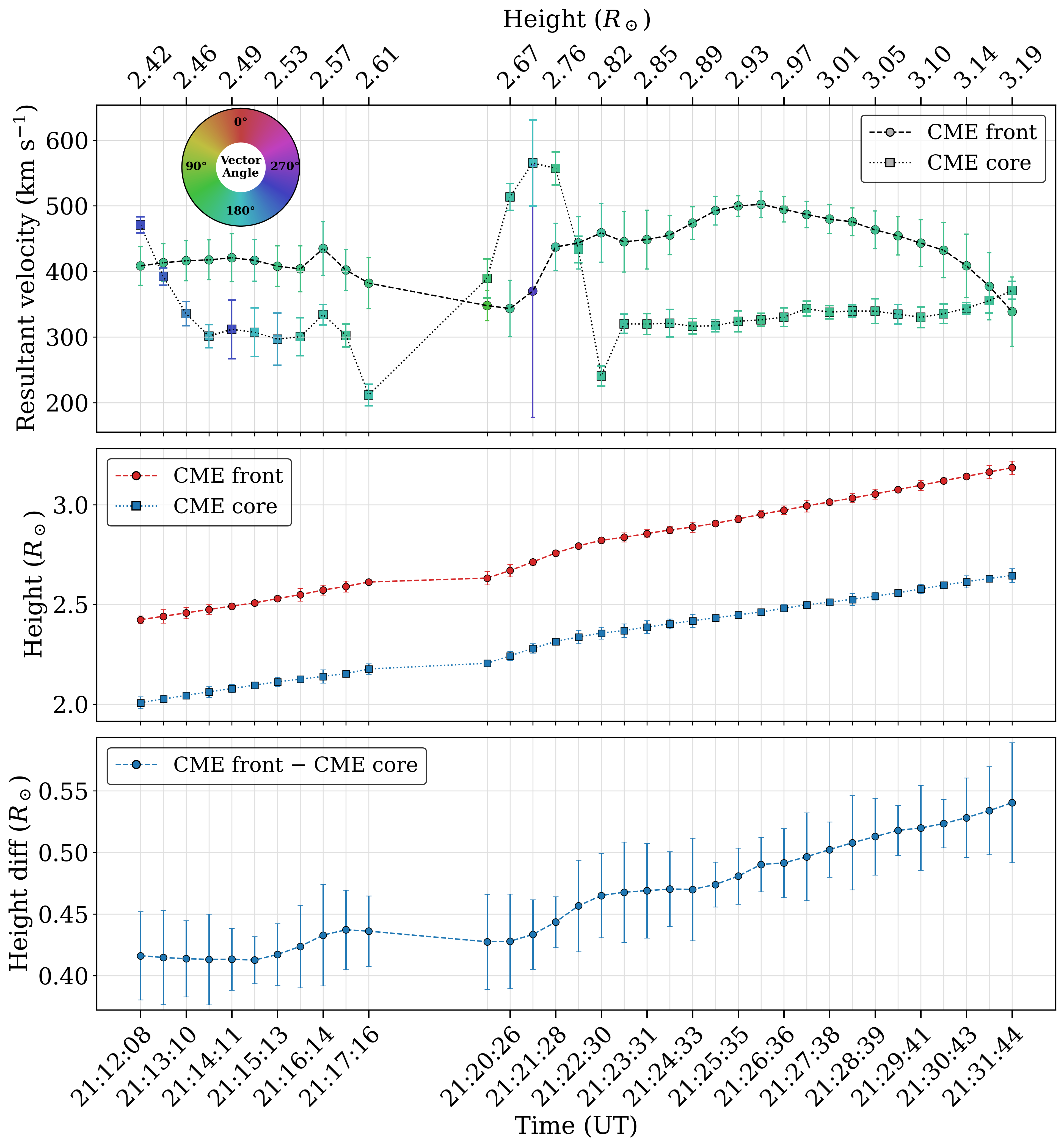}{0.48\textwidth}{(c) PA = 151°}
}

\vspace{-2mm}

\caption{\footnotesize
Substructure kinematic diagnostics for the 2024 March 29 CME observed by METIS at two PAs. Panel~(a) shows co-temporal observations from METIS and FSI, confirming the presence of a prominence-associated core at PA~151°. The structure is clearly detected in the METIS UV and FSI 304 \AA\ passbands, but lacks a corresponding EUV counterpart at PA~136°. The green contours represent structures detected through intensity thresholding, while the CME front and core used for analysis are manually delineated in blue, following features that remain coherent across frames.
}
\label{fig:metis20240329}
\end{figure*}
\begin{figure*}[ht!]
\centering
\gridline{
\fig{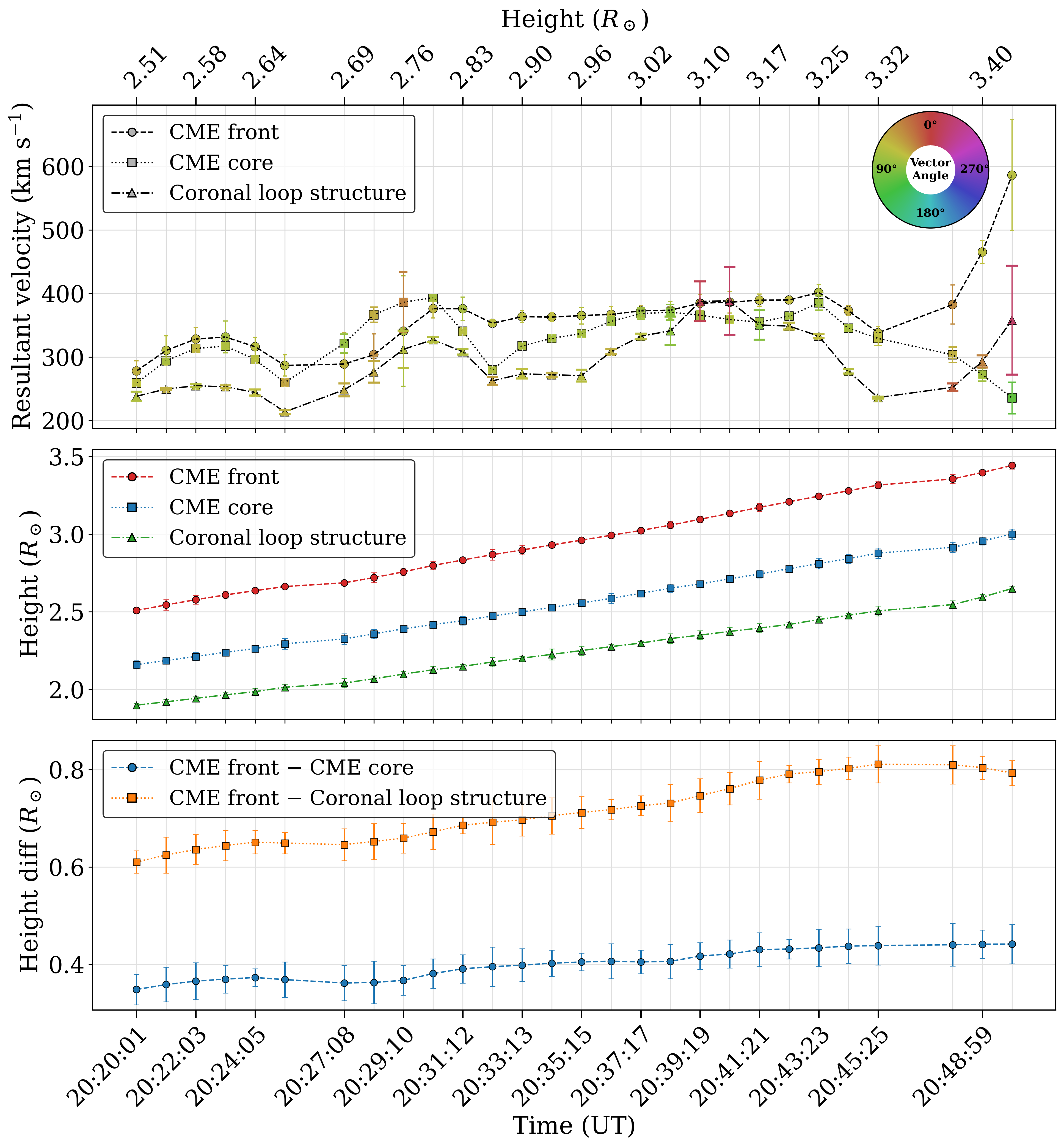}{0.48\textwidth}{(a) PA = 62°}
\fig{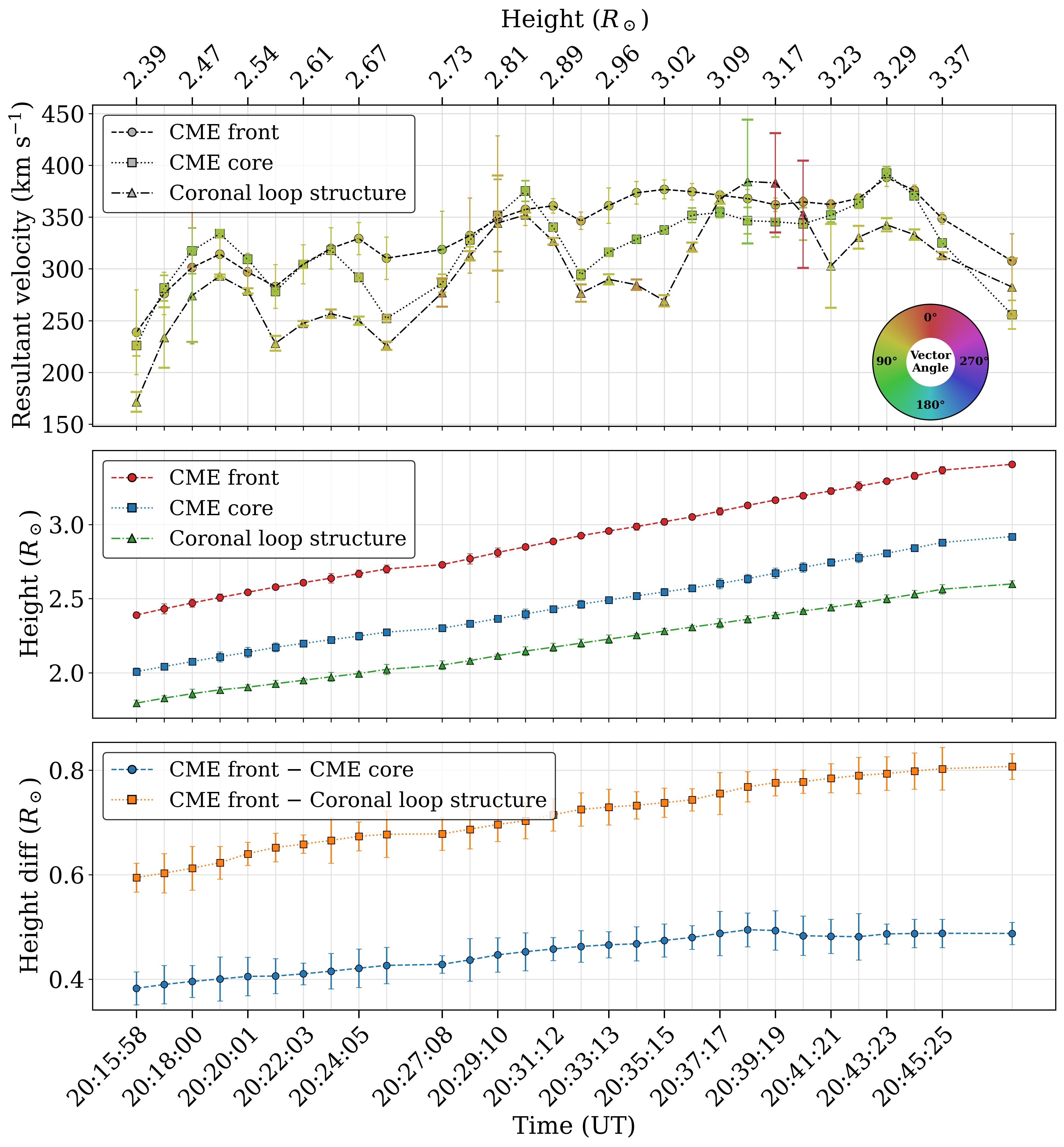}{0.48\textwidth}{(b) PA = 64°}
}
\vspace{-2mm}
\caption{\footnotesize
Substructure kinematic diagnostics for the 2023 October 5 CME observed by METIS at two PAs.
}
\label{fig:metis20231005}
\end{figure*}

To visualise these results, we present a set of multi-panel plots for selected events and position angles. Each figure corresponds to a single CME event observed at different PAs, illustrating the temporal evolution of the resultant velocities, radial heights, and inter-structure height differences for the CME substructures. These examples highlight the spatially dependent kinematic behaviour within CMEs (Figures~\ref{fig:aspiics20250716}, \ref{fig:metis20240329}, and~\ref{fig:metis20231005}).

We examined a CME observed by ASPIICS on 2025 July 16 (See Figure~\ref{fig:xt_vs_flow_validation}a) from three closely spaced position angles (PAs): 280°, 283°, and 285°. The results are shown in Figure~\ref{fig:aspiics20250716}, which presents the time evolution of the resultant flow velocity (top panel), helioprojective height (middle panel), and height difference between the front and core (bottom panel) for each PA.

All three PAs exhibit a clear impulsive acceleration phase in both the front and the core simultaneously, peaking around 17:31:07~UT, when the CME front reaches approximately $2.20~R_\odot$, followed by a decay phase. This phase marks the onset of structural divergence between the front and core. At PA~280° (Figure~\ref{fig:aspiics20250716}a), the core's height begins to lag noticeably behind the front shortly after 17:34:07~UT ($\sim2.32~R_\odot$), with the height offset increasing from $\sim$0.21 to $0.45~R_\odot$ over the next 15 minutes. During this interval, the front velocity increases toward 410~$\mathrm{km\ s^{-1}}$ while the core velocity declines, reflecting a redistribution of momentum from the trailing core to the leading edge.

At PA~283° (Figure~\ref{fig:aspiics20250716}b), following a sharp drop in core speed at 17:33:39~UT, the front reaches its local minimum velocity at 17:34:37~UT ($\sim$58~seconds delay). This indicates that both components begin decelerating nearly simultaneously, but the core attains its minimum earlier than the front. After reaching its local minimum, the core begins to accelerate earlier and attains its post-dip maximum at 17:35:07~UT, whereas the front reaches its maximum later at 17:37:07~UT ($\sim$120~seconds delay). While the onset of deceleration is nearly simultaneous, the out-of-phase attainment of both minimum and recovery phases suggests a period of momentum exchange accompanied by a subsequent increase in their height difference. The core eventually settles into a slower regime around 280–300~$\mathrm{km\ s^{-1}}$, while the height offset continues to grow, reaching $0.48~R_\odot$.

PA~285° (Figure~\ref{fig:aspiics20250716}c) exhibits similar dynamics to PA~283°, but with a larger temporal offset. Both components begin decelerating nearly simultaneously, but the core reaches its minimum at 17:34:07~UT, while the front attains its minimum at 17:35:37~UT ($\sim$90~seconds delay). Thereafter, the core begins to accelerate earlier and reaches its post-dip maximum at 17:35:07~UT, whereas the front peaks later at 17:37:37~UT ($\sim$150~seconds delay). Across all PAs, the height dispersion between the front and the core increases substantially after the initial impulsive acceleration decay phase and once the front exceeds $2.32~R_\odot$, signifying the transition from a coupled evolution of the front and core to a regime of rapid radial expansion.

We next analyzed a CME observed by METIS on 2024 March 29 (see Figure~\ref{fig:metis20240329}a) at two distinct PAs: 136° and 151°. Figure~\ref{fig:metis20240329}b and~\ref{fig:metis20240329}c summarizes the kinematic evolution of the CME front and core at both PAs. In both cases, the height and velocity profiles reveal progressive divergence between the two structures, but the nature and timing of this decoupling differ significantly.

At PA~136° (Figure~\ref{fig:metis20240329}a), where no clear EUV counterpart of the core was identified in FSI or METIS UV channels, the velocity profiles exhibit pronounced oscillatory behaviour. Between 21:27:38 and 21:34:18~UT, the front and core exhibit weakly out-of-phase oscillations. A comparison of corresponding crests and troughs shows that the core generally lags the front by $\sim$70~s on average, with a standard deviation of $\sim$50~s. This variability indicates the absence of a consistent phase relationship and suggests that the front responds (i.e., accelerates or decelerates) earlier than the core during this interval. For instance, while the front velocity had begun to increase slightly, the core reached a local minimum of $\sim$225~$\mathrm{km\ s^{-1}}$ at 21:30:12~UT. This phase-shifted motion reflects dynamic coupling between the two structures. The height dispersion begins its most sustained increase at 21:33:17~UT, when the front reaches $\sim2.95~R_\odot$, and grows steadily from $\sim$0.35 to 0.39~$R_\odot$.

In contrast, PA~151° (see Figure~\ref{fig:metis20240329}b) features a core structure associated with a prominence eruption, confirmed by co-temporal signatures in METIS UV and FSI 304~\AA\ images (Figure~\ref{fig:metis20240329}a). Here, the dynamics are markedly more impulsive. The core velocity surges to a peak of $\sim$565~$\mathrm{km\ s^{-1}}$ at 21:20:26~UT—briefly outpacing the front, which moves at only $\sim$345~$\mathrm{km\ s^{-1}}$ at that moment. Following this sharp rise, the core velocity declines and stabilizes near 330~$\mathrm{km\ s^{-1}}$. The height offset starts increasing rapidly soon after the core's impulsive acceleration, with a clear inflection point at 21:20:57~UT, when the front reaches $\sim2.76~R_\odot$. This relatively earlier and pronounced separation suggests internal force imbalances, potentially driven by the erupting prominence material.

The CME observed by METIS on 2023 October 5 (see Figure~\ref{fig:xt_vs_flow_validation}b) exhibited relatively coherent propagation across its substructures. At both PA~62° and PA~64°, three features were tracked: the CME front, the core, and an intermediate loop-like structure (Figure~\ref{fig:metis20231005}). Unlike the earlier events, this case shows stronger temporal coupling among the substructures. The resultant velocity profiles reveal that the front, core, and loop evolve largely in phase, with well-aligned crests and troughs throughout the time series. The radial height separations between these features increase gradually, but do not show sharp transitions or inflection points. These relatively constant offsets and kinematic synchrony suggest a quasi-rigid expansion regime, where the substructures maintain similar velocity evolution and exhibit a slow and minimal increase in their height separation.

\begin{figure*}[ht!]
\centering
\gridline{
  \fig{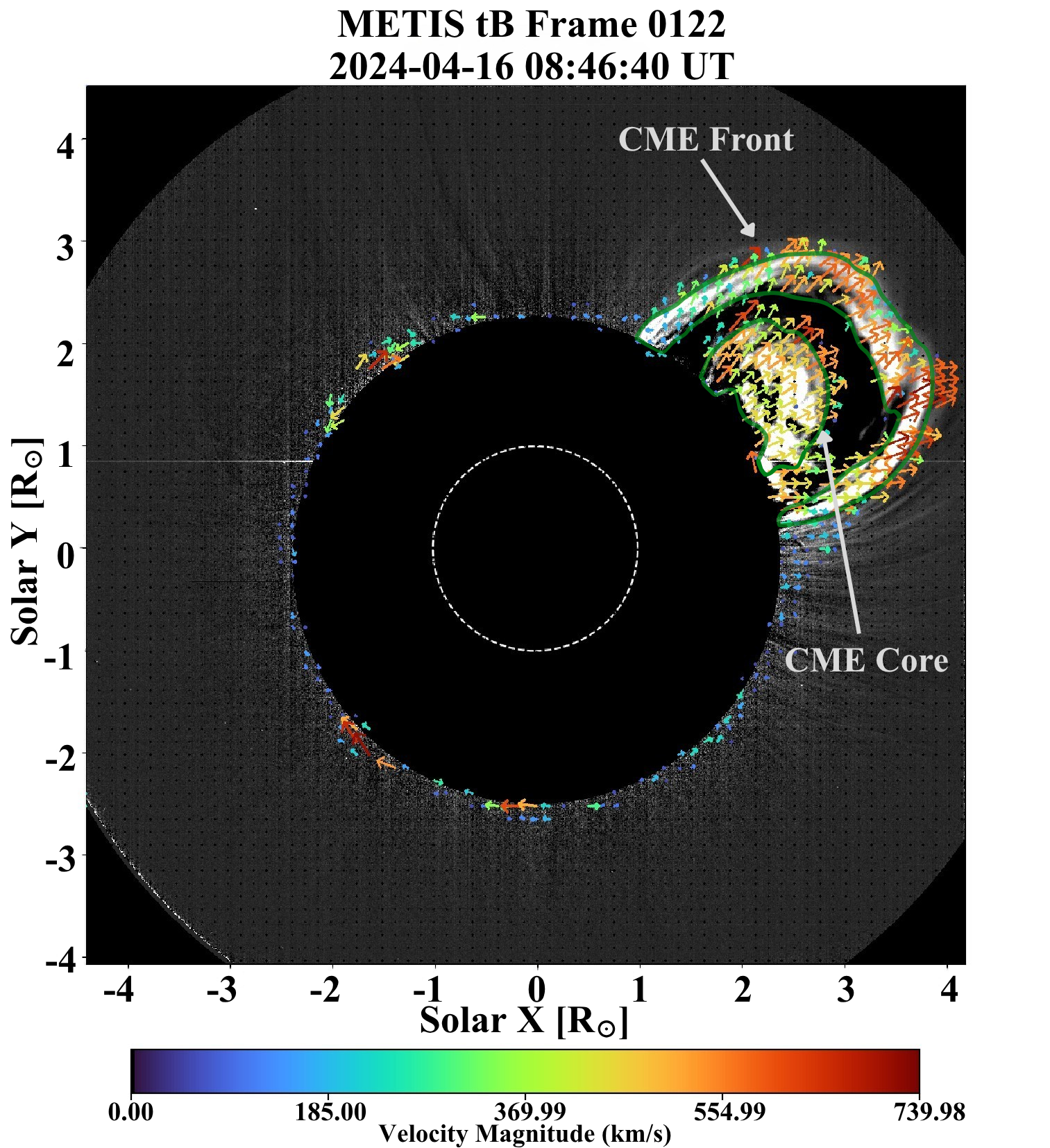}{0.45\textwidth}{(a) Optical flow map of the CME event}
  \fig{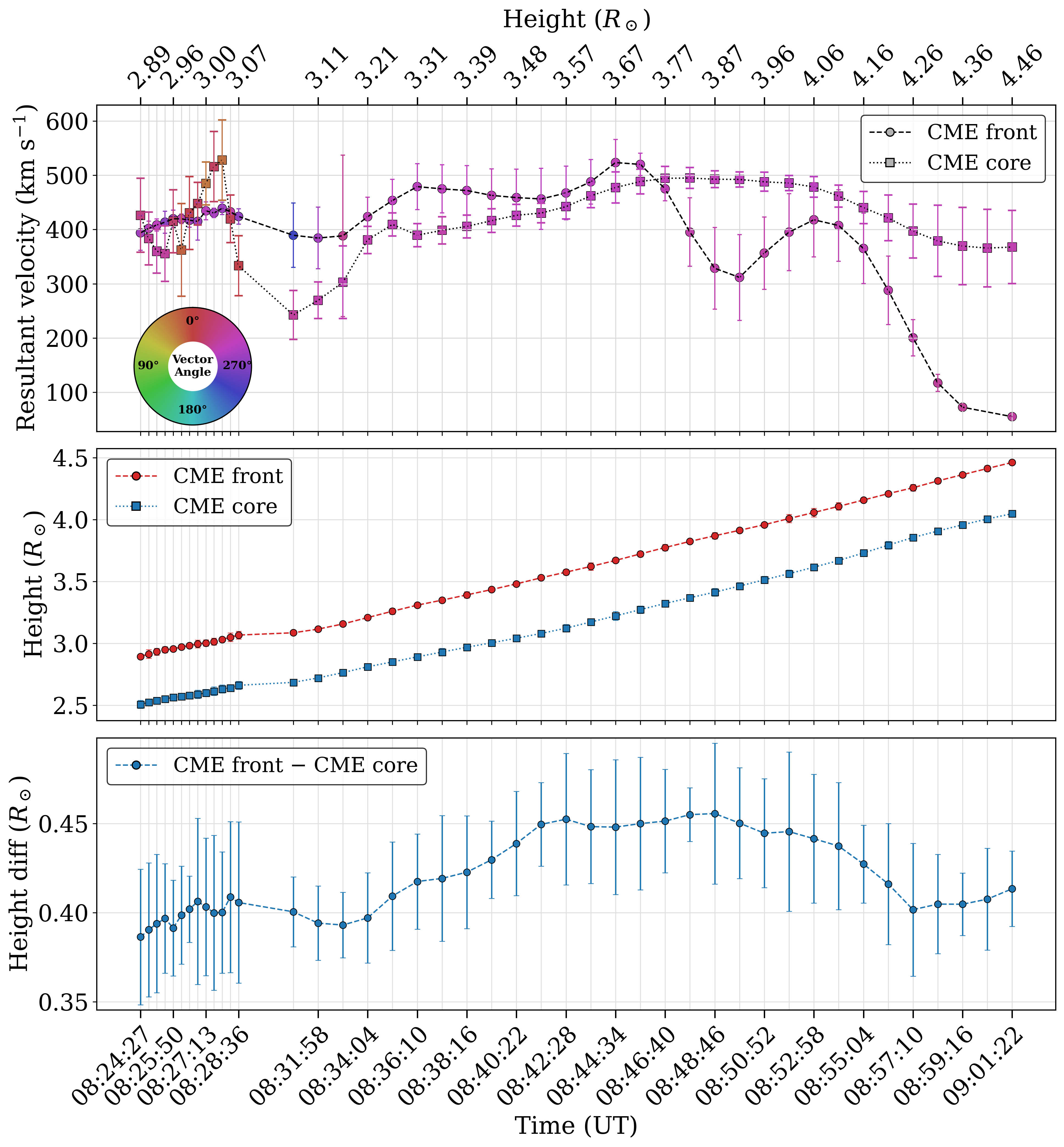}{0.48\textwidth}{(b) PA = 305°}
}
\vspace{-2mm}
\caption{\footnotesize
Substructure kinematic diagnostics for the 2024 April 16 CME at PA = 305°. Panel~(a) shows an optical flow frame highlighting substructure motion, while panel~(b) presents the resultant flow velocity, height evolution, and internal height offset between the CME front and core.
}
\label{fig:metis20240416}
\end{figure*}

The 2024 April 16 CME observed by METIS at PA~305° exhibits an unusual kinematic profile (Figure~\ref{fig:metis20240416}). While the front maintains a relatively steady propagation speed, the core undergoes a marked impulsive acceleration peaking at 08:27:53 UT, and eventually surpasses the front in velocity at 08:46:40 UT. This dynamic leads to a decrease in the height offset—from $\sim$0.45~$R_\odot$ to $\sim$0.39~$R_\odot$. In this case, the CME core's distance from the front decreases, similar to that reported in a previous event by \citep{Song2025}.


\subsection{Intra-structural Velocity Dispersion within CME}
\label{sec:Inrta-dispersion}

\begin{figure*}[ht!]
\centering
\gridline{
\fig{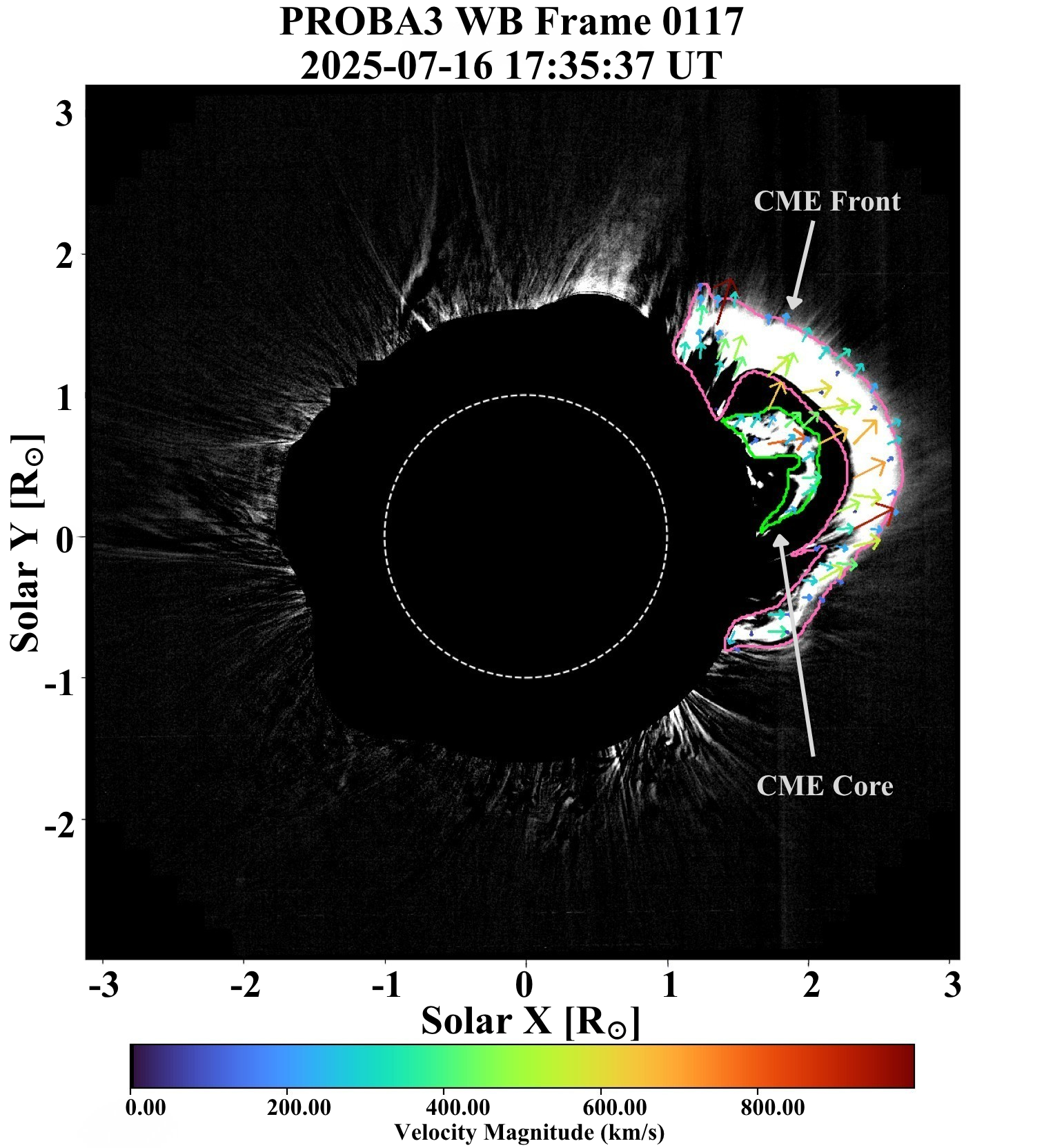}{0.40\textwidth}{(a) Optical flow map with lasso selection of front and core}
\fig{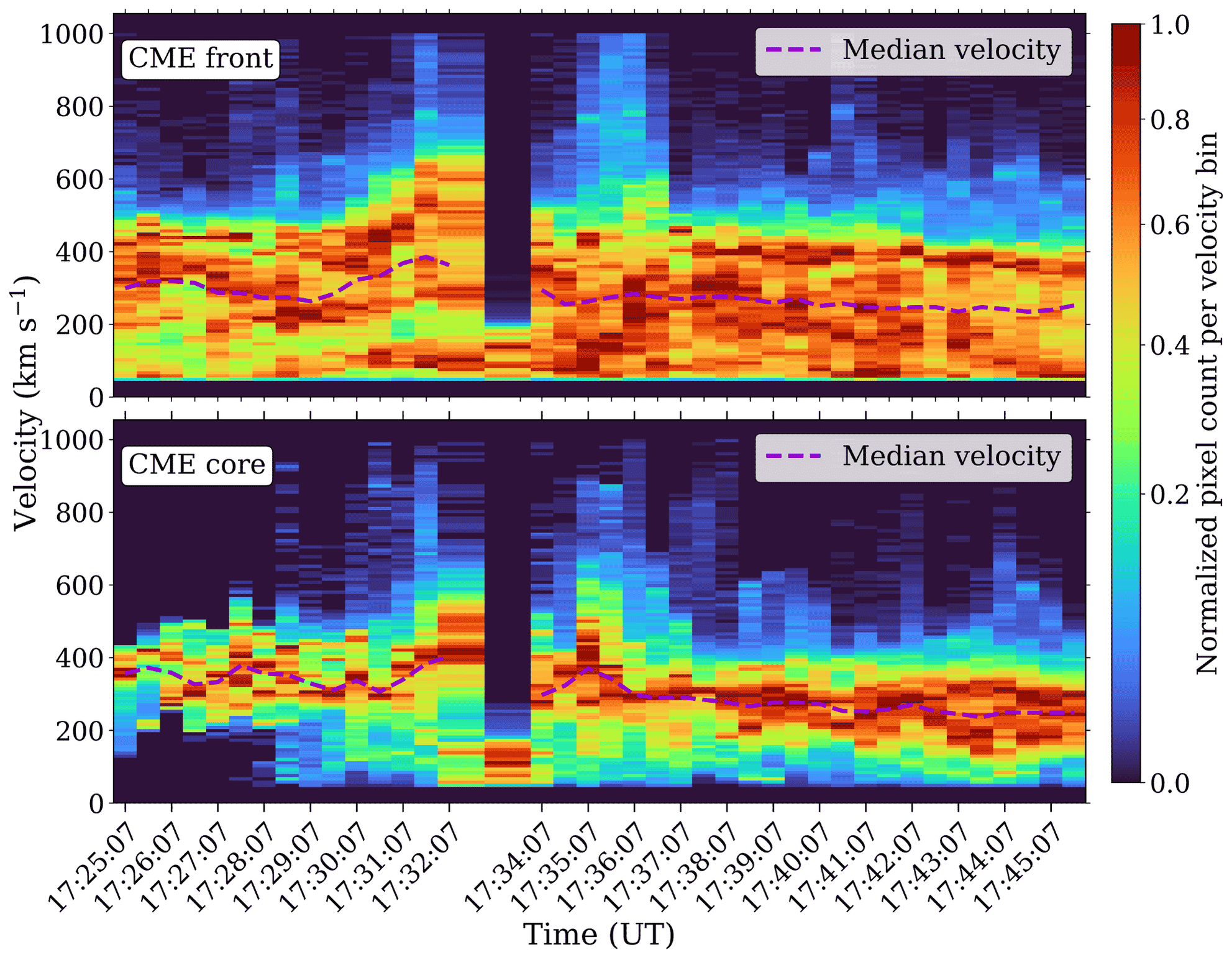}{0.50\textwidth}{(b) Velocity dispersion in the front and core}
}
\vspace{-2mm}
\gridline{
\fig{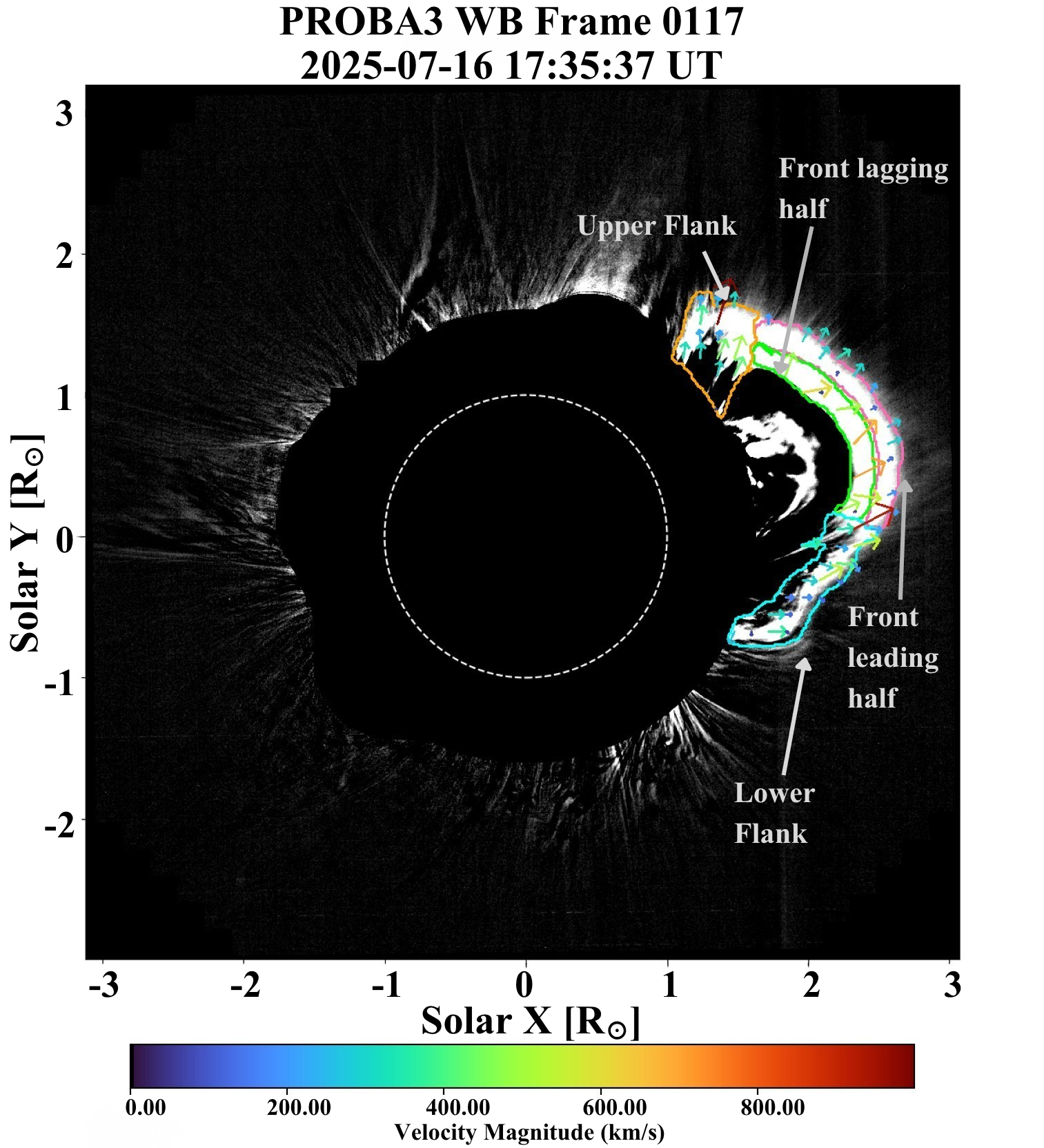}{0.40\textwidth}{(c) Optical flow map with lasso selection of front subregions}
\fig{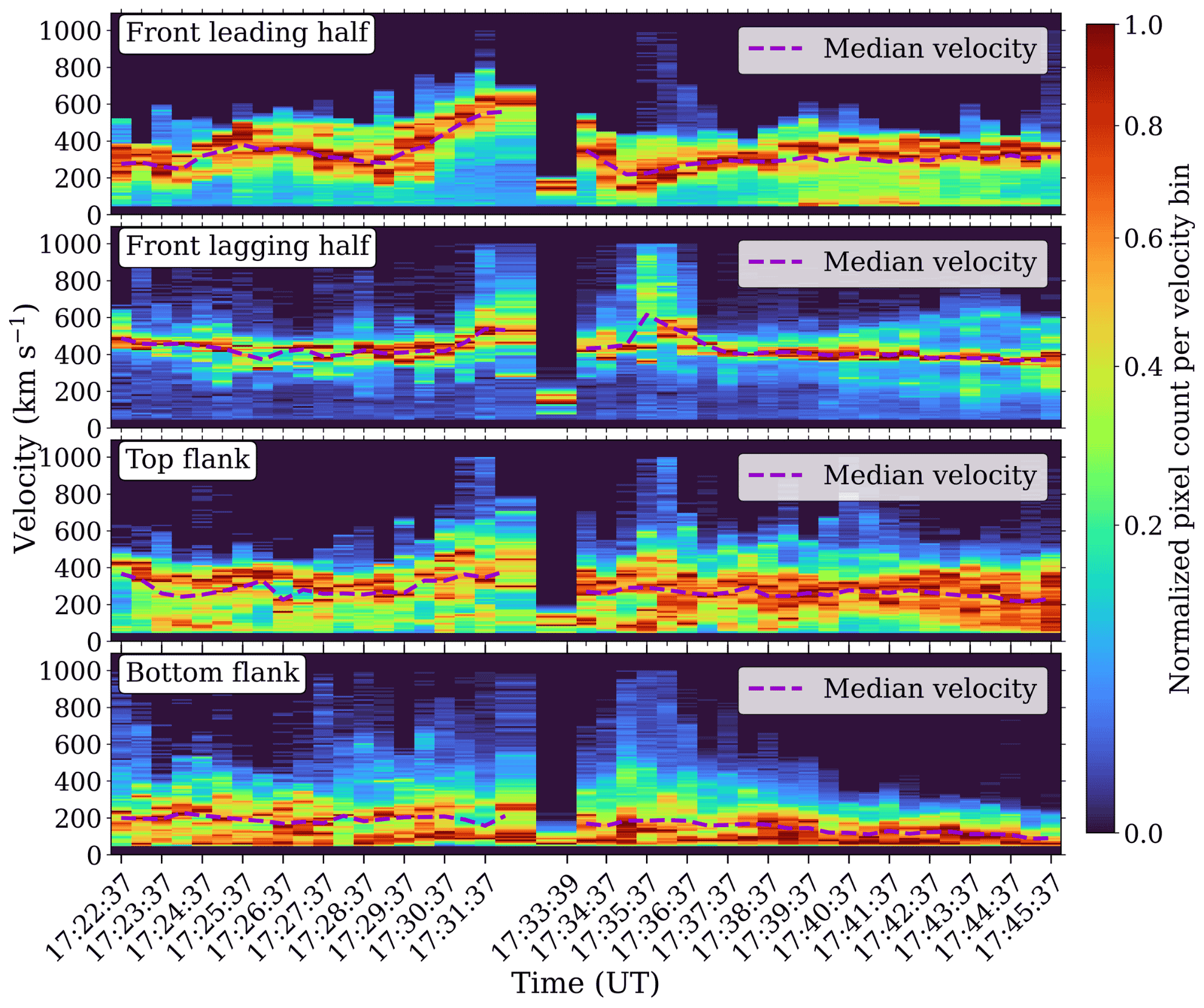}{0.50\textwidth}{(d) Velocity dispersion across front subregions}
}
\vspace{-2mm}
\caption{\footnotesize
Internal velocity dispersion within the 2025 July 16 CME observed by ASPIICS. Panel (a) shows the corresponding lasso-based manual selections used for the front and core. Panel (b) presents the corresponding velocity distributions for these regions, where each column represents a velocity histogram at the given time step, and the colorbar denotes the normalised velocity density. Panel (c) displays the manually selected regions within the front, overlaid on the optical flow velocity map. Panel (d) presents the velocity distributions across four subregions of the front. The dark violet dashed curve indicates the median velocity, calculated only when there are sufficient velocity values.
}
\label{fig:velocity_dispersion_regions}
\end{figure*}

We now shift focus from the comparison of kinematics between CME substructures discussed in the previous subsection to the velocity dispersion within individual substructures. While the earlier analysis focused on the relative evolution of the front and core across different PAs, here we examine the velocity spread and distribution within these structures. We use the 2025 July 16 ASPIICS event as a case study, owing to its clearly observed impulsive acceleration in both the front and core during its propagation through the middle corona (Figure~\ref{fig:aspiics20250716}). This event is particularly relevant because the impulsive acceleration is known to occur predominantly in the middle corona \citep{West2023}. To identify which sub-regions within these substructures contribute to this acceleration, we applied optical flow tracking and used lasso-based manual selection to isolate coherent features across consecutive frames (Figure~\ref{fig:velocity_dispersion_regions}a, c).

Figure~\ref{fig:velocity_dispersion_regions}b presents the velocity distributions for the CME front and core. The core shows a relatively narrow distribution, with most velocities concentrated near the median, suggesting coherent bulk motion. In contrast, the front displays a broader distribution spanning low to high velocities, indicating a more heterogeneous internal kinematics. Both structures exhibit a sharp rise in their median velocities around 17:31:07~UT corresponding to the impulsive acceleration phase identified in Figure~\ref{fig:aspiics20250716}.

To better understand the internal complexity of the CME front and identify which regions contribute to impulsive acceleration versus lower-speed flows, we divided it into four spatial sectors: the front leading half, front lagging half, upper flank, and lower flank (Figure~\ref{fig:velocity_dispersion_regions}d). The leading half exhibited a pronounced rise in median velocity along with high-velocity features, consistent with impulsive acceleration. The lagging half maintained a relatively steady median but also included a population of high-speed vectors. However, the main contributors to the low-velocity population were the flanks, particularly the lower flank. The lower flank showed a narrow distribution, concentrated at lower velocities. In contrast, the upper flank displayed a wider spread, with contributions from both low and high velocities. This contrast suggests that even within the lateral extents of the front, the CME exhibited pronounced internal variability. Such spatial variability not only quantifies the distinct contributions of different regions within the CME front to the overall velocity distribution but also highlights the capability of the optical flow method to resolve such internal kinematic structure. While the front exhibited a broad velocity distribution, the core showed a comparatively narrow spread, with values clustered close to its median, indicating more uniform evolution. The impulsive acceleration within the front is also not spatially uniform, being primarily driven by the leading half. Consequently, most of the internal velocity dispersion is concentrated in the front, which conventional leading-edge tracking approaches cannot capture.

\subsection{Evaluating Radial Self-Similar Expansion in CME Front and Core}
\label{sec:self-similarity}

Self-similar expansion describes a regime in which the radial speed of any CME feature scales linearly with its heliocentric distance, i.e., $v \propto r$ \citep{Wood2016}, implying that the overall shape of the CME is preserved during propagation. While \citet{Maričić2009} showed that the whole magnetic arcade of erupting CMEs can exhibit self-similar characteristics, subsequent work has revealed a more detailed picture. \citet{Cremades2020} and \citet{Majumdar2020} found that CMEs
expand non-self-similarly in the inner corona, whereas \citet{Subramanian2014} showed that the expansion is roughly
self-similar in the outer corona. Crucially, both studies evaluated the overall CME structure rather than the expansion of distinct internal substructures.

To examine whether the CME front and core expand in a self-similar manner during propagation, we adopt the expansion
coefficient $S_r$ formulated by \citet{Wood2016}:

\begin{equation}
    S_r = \frac{(r_1 + r_2)(v_1 - v_2)}{(r_1 - r_2)(v_1 + v_2)},
    \label{eq:sr}
\end{equation}

\noindent where $r_1$ and $r_2$ are the heliocentric heights of the leading edge (LE) and trailing edge (TE) of the feature under consideration, and $v_1$ and $v_2$ are the corresponding resultant velocities, averaged over the LE and TE regions respectively, at the same position angle and time. We apply Equation~\ref{eq:sr} separately to the CME front and core: for each substructure, we track its LE and TE, divide the substructure into two regions bounded by these edges, and compute $r_1$, $v_1$ and $r_2$, $v_2$ as the heights and averaged velocities of these two regions. This yields two independent sets of self-similar expansion coefficient, $S_r^{\text{front}}$ and $S_r^{\text{core}}$. We assume radial propagation for both substructures, which is justified by the average resultant velocity angles being closely aligned with the selected position angle in all four events (see the hue wheel insets in Figures~5--8).

The value of $S_r$ can be interpreted physically as follows. When $S_r = 1$, the expansion rate exactly scales with heliocentric distance, corresponding to perfect self-similar expansion. When $S_r > 1$, the substructure expands faster than the self-similar expansion, indicating that the separation between the LE and TE grows more rapidly than is suggestive of self-similar expansion. When $0 < S_r < 1$, the substructure expands more slowly than self-similar expansion. When $S_r = 0$, both LE and TE move at the same speed ($v_1 = v_2$), and there is no relative expansion between them. When $S_r < 0$, the TE velocity exceeds the LE velocity ($v_2 > v_1$), meaning the substructure is contracting.

\begin{figure*}[ht!]
\centering
\gridline{
\fig{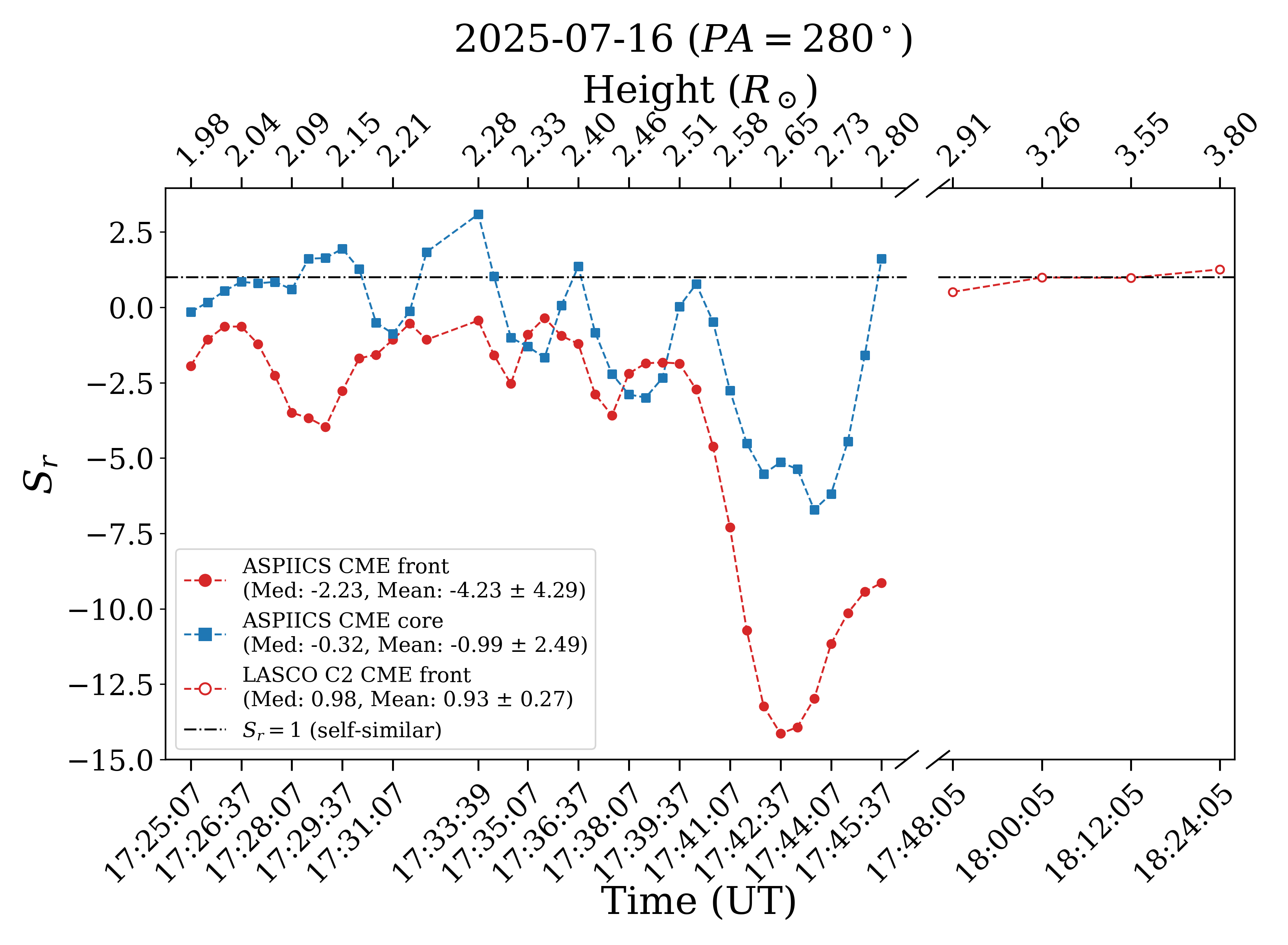}{0.49\textwidth}{(a)}
\fig{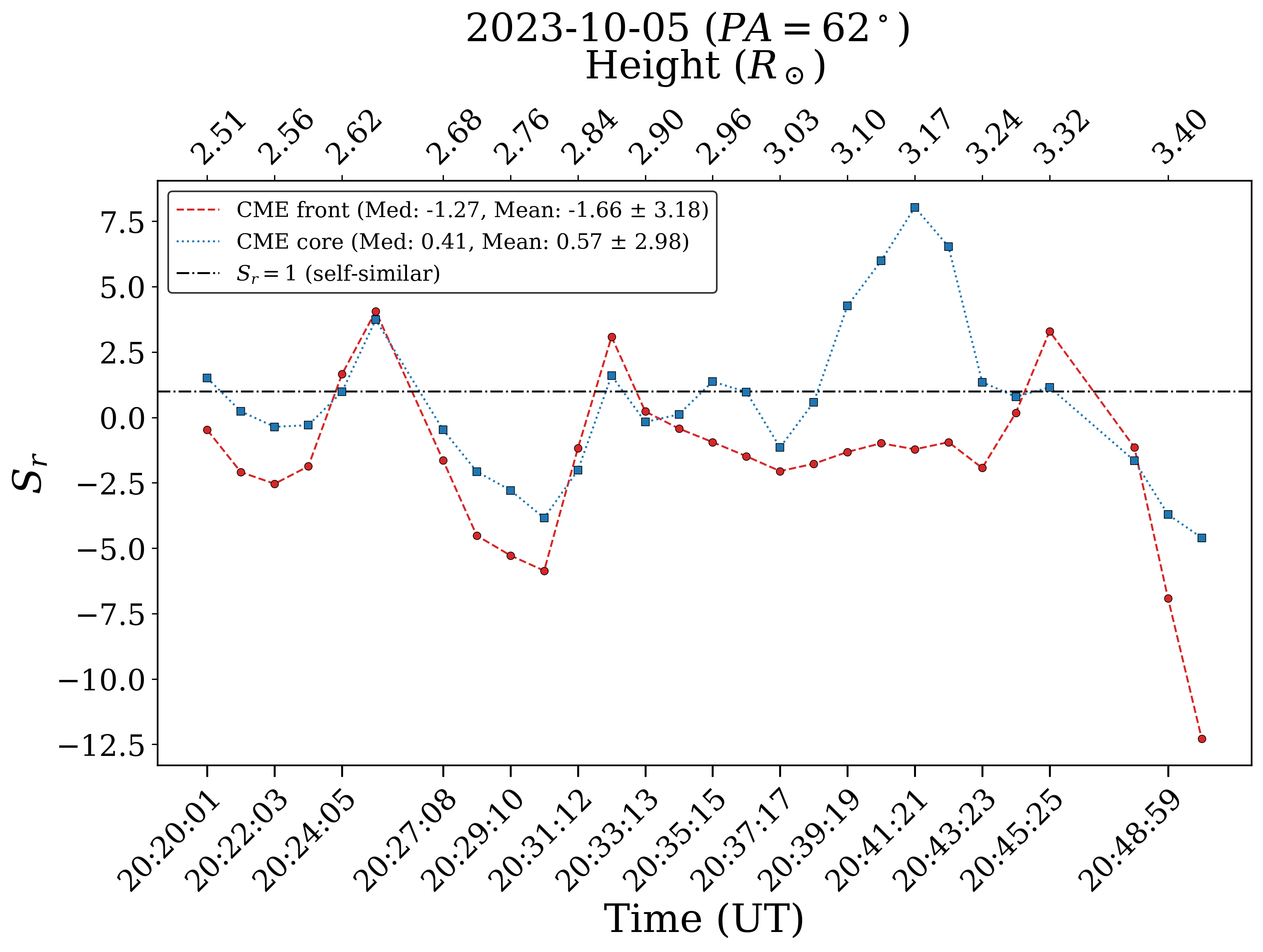}{0.49\textwidth}{(b)}
}
\vspace{-4mm}
\gridline{
\fig{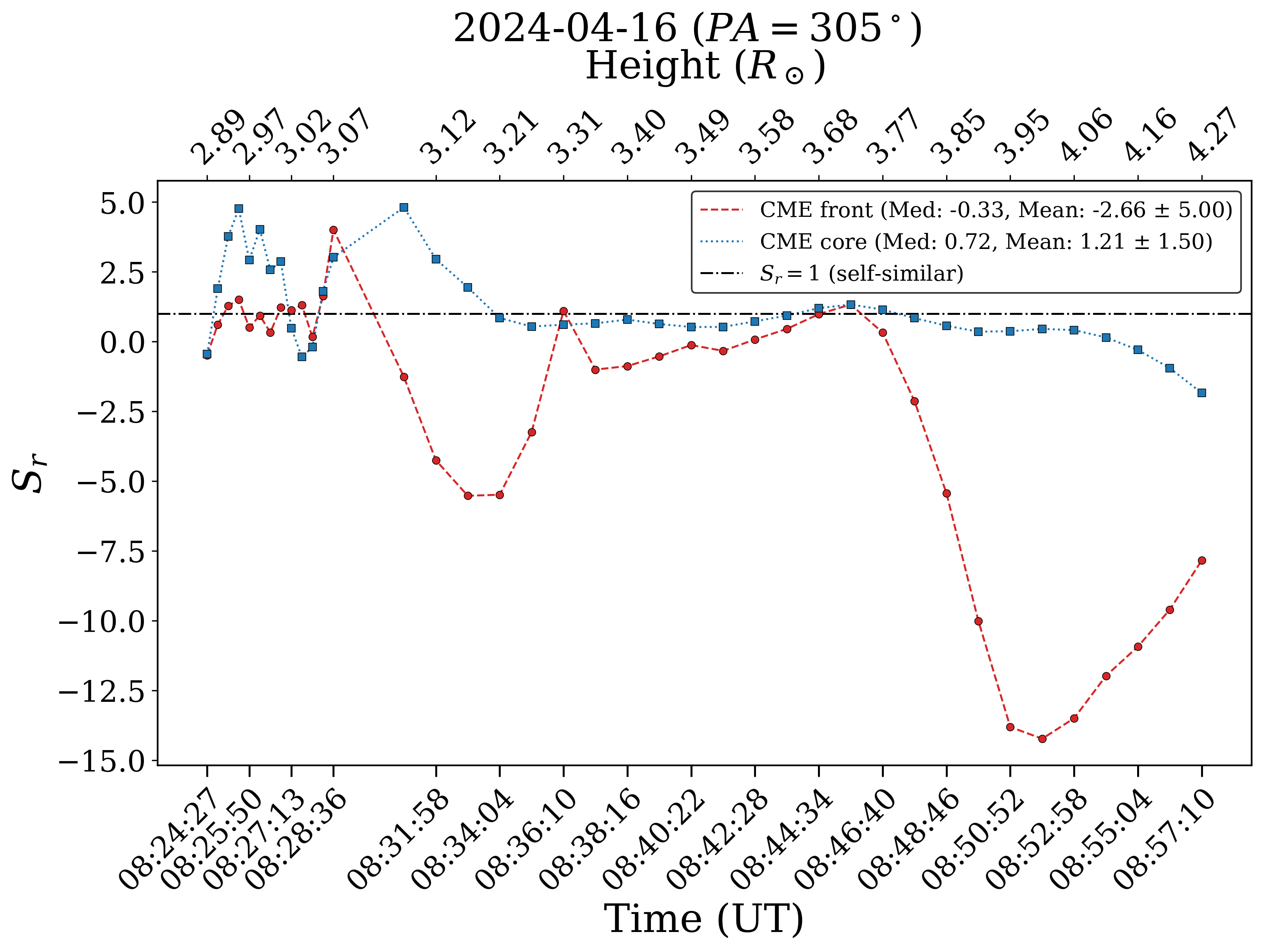}{0.49\textwidth}{(c)}
\fig{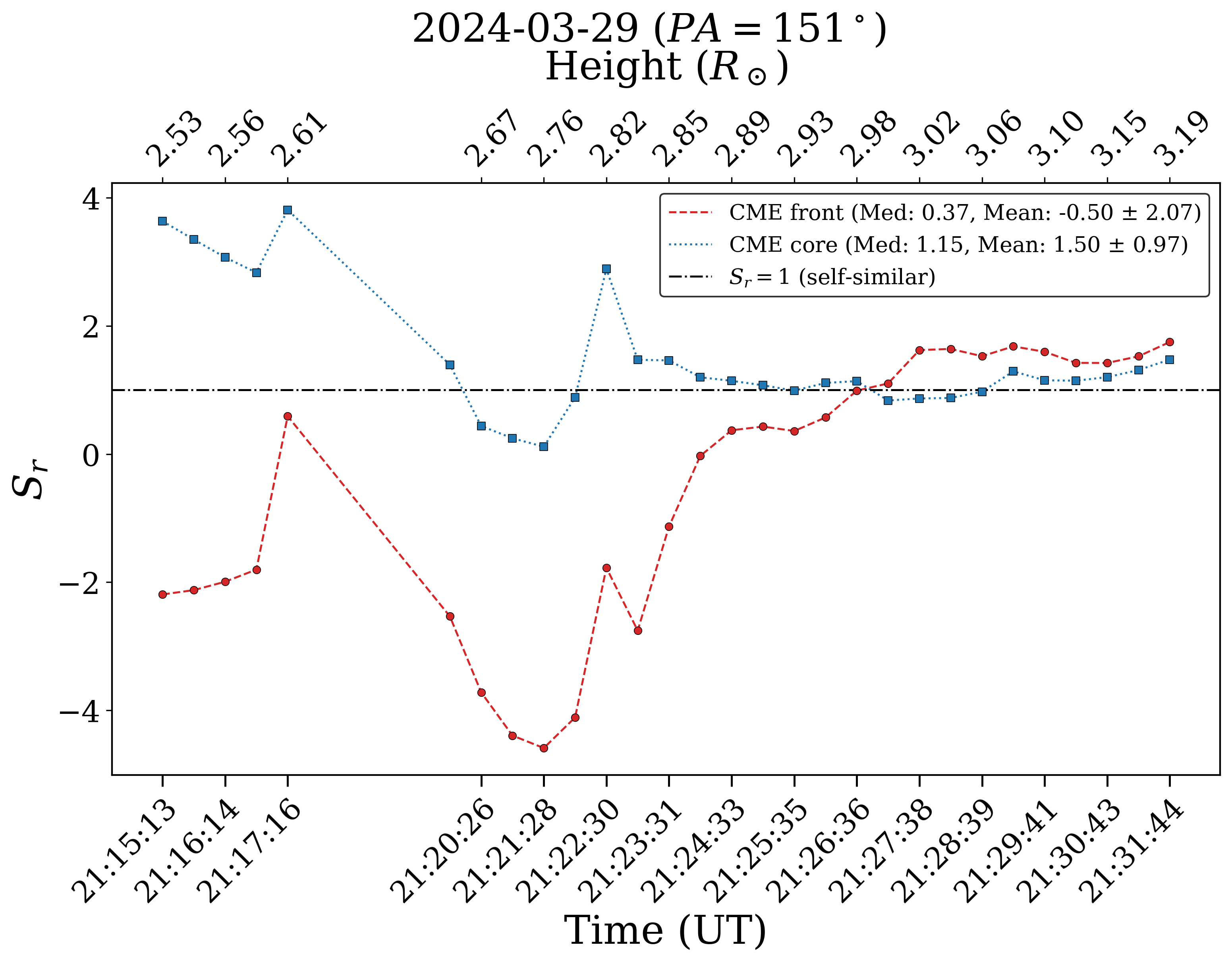}{0.49\textwidth}{(d)}
}
\vspace{-4mm}
\gridline{
\fig{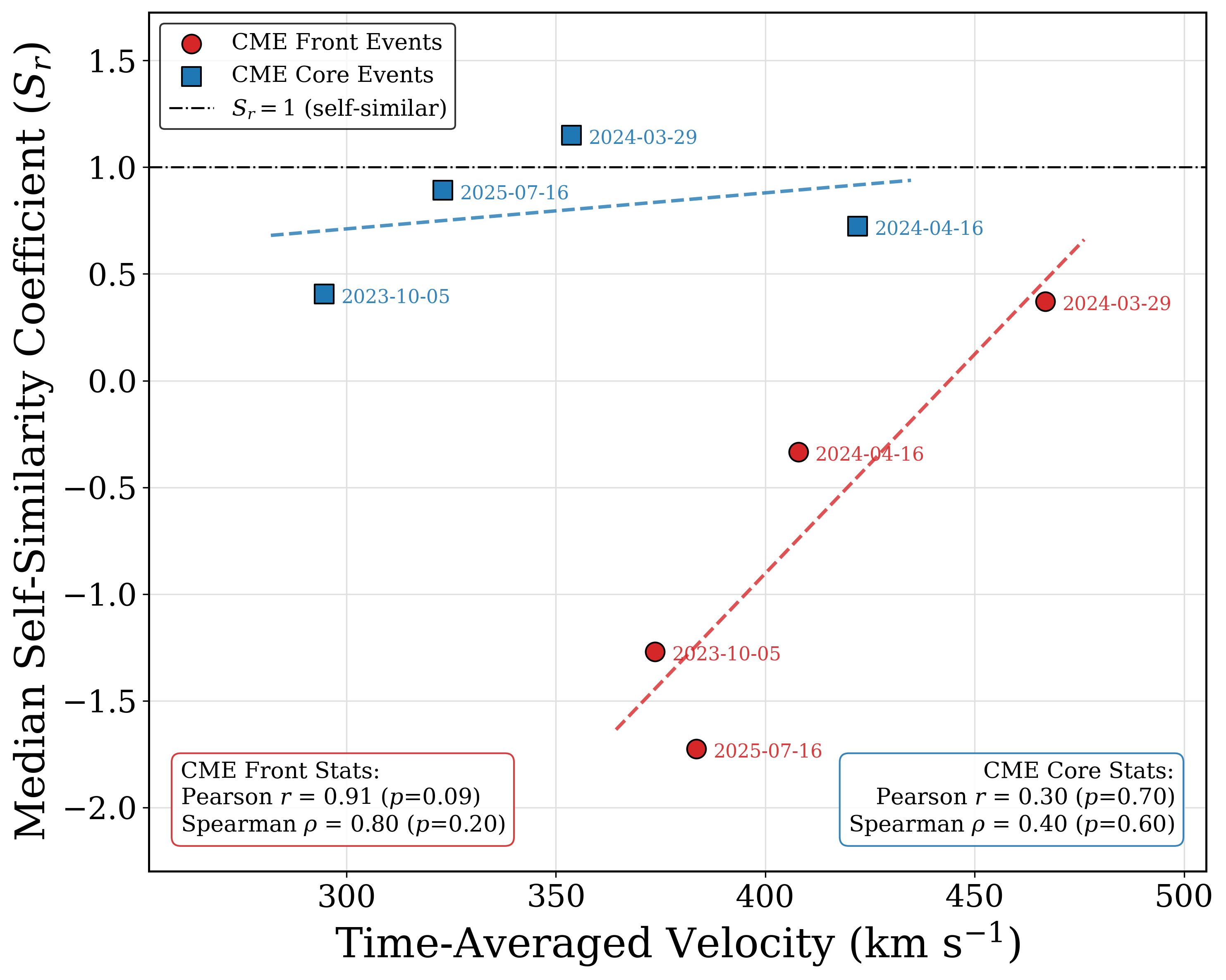}{0.48\textwidth}{(e)}
\fig{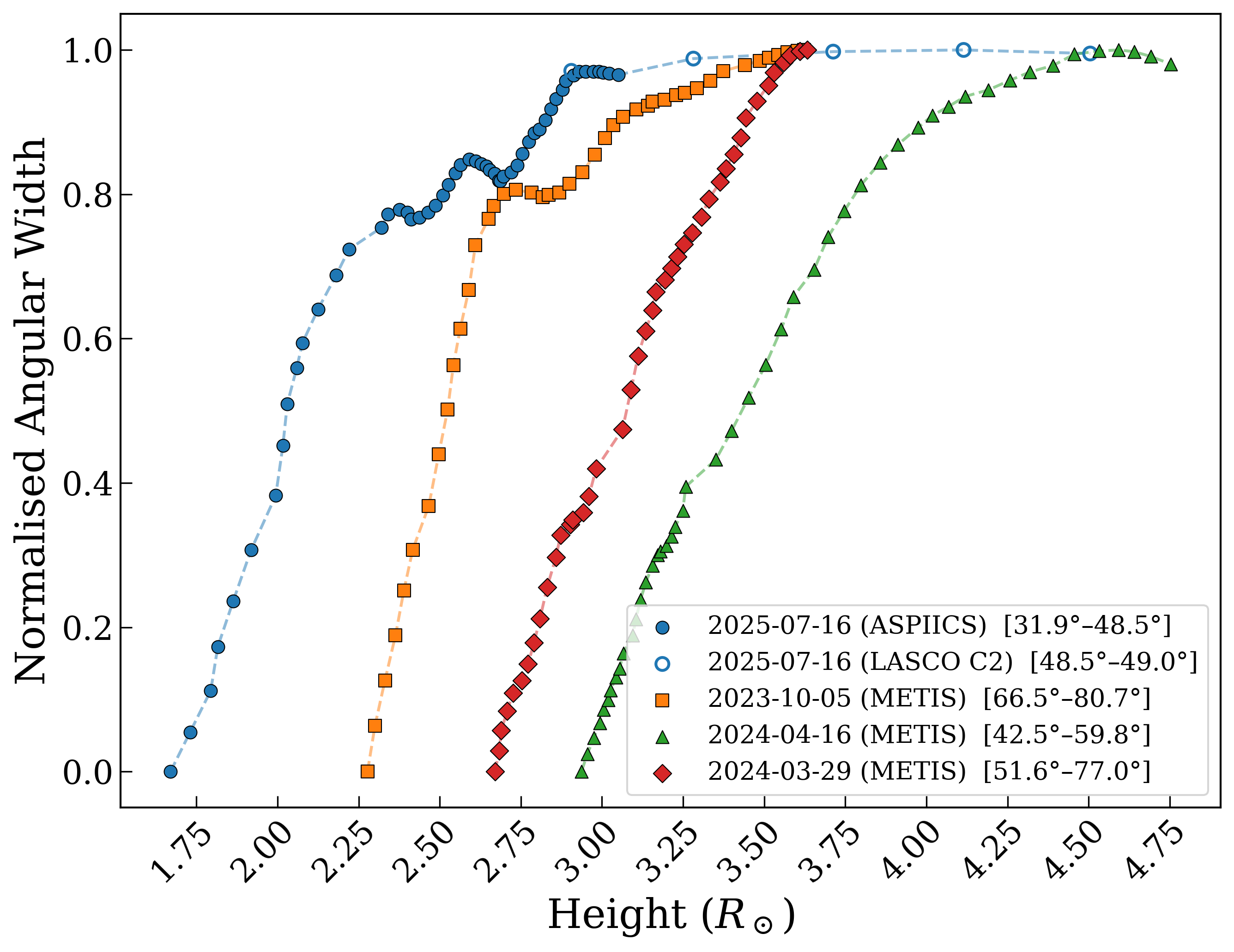}{0.49\textwidth}{(f)}
}
\caption{(a–d) Frame-by-frame self-similarity coefficient $S_r$ as a function of CME front height for the front (red) and core (blue), shown separately for all four events. In panel (a), the time and height axis break indicates a different scaling used for the LASCO C2 data. (e) Median $S_r$ versus time-averaged resultant velocity for the CME front and core across all four events. (f) Normalised CME angular width versus the front height.}
\label{fig:self_similarity_expansion}
\end{figure*}

Figure~\ref{fig:self_similarity_expansion}(a-d) shows the frame-by-frame $S_r$ as a function of CME front height for all four events.

To address whether the degree of self-similarity varies systematically with height, we examined the $S_r^{\text{front}}$ and $S_r^{\text{core}}$ trends as a function of height in each event. While no single, consistent trend with height is apparent across all four events, two recurring features emerge: the front frequently recovers toward more self-similar values ($S_r \sim 1$) at larger heights following an episode of strong contraction in the lower heights, whereas the core remains comparatively close to self-similar or super-self-similar values throughout the tracked height range in most events.

Across all four events, the median $S_r^{\text{core}}$ is close to or above 1 in three of the four events (0.41, 0.72, 1.15), while the median $S_r^{\text{front}}$ remains negative or near zero in all four events (-2.23, -1.27, -0.33, 0.37). This systematic offset of the degree of self-similarity between the front and core may signal compression of the CME front in the inner corona, as it interacts with the ambient corona while being trailed by the core.

We further find that the front exhibits substantially greater variability than the core: in every event, the mean $S_r^{\text{front}}$ deviates markedly from its median. Notably, the front shows a recovery toward self-similarity at higher heights in three of the four events (2025-07-16, 2024-04-16, and 2024-03-29), which we discuss in detail below.

In the 2025-07-16 event (Figure~\ref{fig:self_similarity_expansion}a), the ASPIICS front $S_r$ plunges to $\sim-13$ near $2.68~R_\odot$, after which it begins increasing again to $\sim-10$ by the edge of the ASPIICS field of view. To probe this trend at larger heights, we corroborated the ASPIICS measurements with LASCO~C2 observations, which share the same Sun--Earth line of sight. We tracked only the CME front in LASCO~C2, since the core appeared substantially deformed in this instrument's field of view. The LASCO~C2 front $S_r$ was found to lie between $0.6$ and $1.3$ across its field of view, consistent with near self-similar expansion. This suggests that the front experiences contraction in the lower coronal heights (ASPIICS observations), and as it propagates through the middle coronal heights (LASCO C2 observations), it attains self-similar expansion. We note, however, that LASCO~C2 has a 12-minute cadence, and, as shown in Section~\ref{sec:ImagePreProc}, running-difference imaging with such a long interval artificially broadens CME structures. This should be kept in mind when interpreting the LASCO~C2 $S_r^{\text{front}}$ values.

In the 2024-04-16 event (Figure~\ref{fig:self_similarity_expansion}c), the front $S_r$ similarly dips to $\sim-5$ near $3.16~R_\odot$, briefly recovers to $\sim1$ near $3.68~R_\odot$, then plunges again to $\sim-14$ near $4~R_\odot$ before showing renewed signs of increasing $S_r$. This recovery is therefore present but not monotonic. The core, by contrast, remained close to self-similar or super-self-similar throughout. In the 2024-03-29 event (Figure~\ref{fig:self_similarity_expansion}d) (tracked at position angle $151^\circ$), the front $S_r$ plunges to $\sim-4.6$ near $2.76~R_\odot$ before rising above 1 at heights around $3~R_\odot$. As in the other events, the core remained close to self-similar or super-self-similar throughout, i.e., it expands faster than the front as it propagates.

A possible physical explanation for this front--core asymmetry is that the front and core behave as dynamically distinct entities rather than as a single coherent structure. As the CME front propagates, it is continuously compressed between the ram pressure of the ambient corona ahead and the outward push of the expanding core trailing behind it. Since the ambient coronal density decreases steeply with heliocentric distance, the ram pressure encountered by the front is expected to decline as the CME propagates outward, which may account for the tendency of $S_r^{\text{front}}$ to recover toward self-similar values at larger heights. The core, in turn, may be comparatively shielded from this ram-pressure interaction by the front itself, which could explain its more consistently self-similar behaviour throughout the tracked height range. Our analysis does not include independent measurements of ambient density or ram pressure, and a quantitative test of this picture would require dedicated MHD modelling or multi-instrument density diagnostics, which we defer for future work.

Figure~\ref{fig:self_similarity_expansion}e shows the relationship between time-averaged velocity and median $S_r$ for the front and core across all four events. We report both the median and mean $S_r$ in the legend of each panel of Figure~\ref{fig:self_similarity_expansion}(a-d); however, the mean is sensitive to individual frames in which $(r_1 - r_2 \approx 0)$ or $(v_2 \gg v_1)$, which causes sudden blow-ups toward large negative $S_r$. We therefore regard the \textit{median} as the more robust statistic for characterising each event. The front exhibits a positive correlation (Pearson $r=0.91$, $p=0.09$; Spearman $\rho=0.80$, $p=0.20$), suggesting that faster fronts tend to be associated with a higher degree of self-similarity. A plausible interpretation is that a higher average velocity reflects a more energetic CME, which allows it to more effectively resist the ram pressure of the ambient medium and thereby sustain more self-similar expansion. No comparable trend is seen for the core (Pearson $r=0.30$, $p=0.70$; Spearman $\rho=0.40$, $p=0.60$). However, with only $N=4$ events, these correlations must be flagged as non-significant, and the results should be regarded as illustrative rather than conclusive. Expanding this analysis to a larger CME sample will be necessary to assess whether a systematic relationship exists between average CME speed and the degree of self-similar expansion of its internal structures.

To assess whether this front--core asymmetry in the degree of radial self-similarity is reflected in the overall angular evolution of the CMEs, and to corroborate our $S_r$-based findings against the previously reported expansion of CMEs as a whole \citep{Cremades2020, Majumdar2022}, we independently measured the angular width of each CME as a function of heliocentric height. In each frame, we identified the outermost extents of the two flank boundaries of the CME and computed the angular width subtended between the flanks and the corresponding front height. For the 2025-07-16 event, we used ASPIICS's level-3 WB data (\href{https://www.sidc.be/proba-3/version-03-release-notes}{ASPIICS Version 03 release notes}), since it comprises a High Dynamic Range composite of different exposures giving us an effective FOV of 1.1 to 3.0~R$_\odot$. This was possible because, for this analysis, we were only concerned with the morphological evolution of the CME. This was supplemented with LASCO~C2 to extend the height range beyond the ASPIICS field of view. Finally, the resulting angular width was normalised to the range [0,1] independently for each event.

Figure~\ref{fig:self_similarity_expansion}f shows the normalised angular width as a function of height for all four events. In three of the four events (2025-07-16, 2023-10-05, and 2024-04-16), the angular width increases rapidly at lower heights and progressively slowly at larger heights, eventually saturating near unity, consistent with the non-self-similar expansion of CME angular widths in the inner corona \citep{Cremades2020,Majumdar2020,Majumdar2022}. In the 2024-03-29 event, the angular width approaches unity only at the very edge of the available METIS FOV, so we cannot confirm whether it has genuinely stabilised or would continue to evolve at larger heights. We further note that the height at which the angular width approaches saturation differs across events (2.9 to 4.5~R$_\odot$). A larger event sample, together with a more systematic characterisation of this transition, is needed to assess whether the deceleration in angular width growth is generally linked to the front--core radial self-similarity trends discussed above.


\section{Discussion and Conclusion}
\label{sec:Discussion}

In this study, we developed and applied an optical flow-based framework (DOFCAT) to quantify CME kinematics and internal velocity distributions using high-resolution coronagraph observations from ASPIICS/Proba-3 and METIS/Solar Orbiter. The framework includes data preprocessing to generate running-difference image sequences with optimized differencing intervals, along with post-processing to enhance CME features and suppress noise. These sequences were then used to derive dense, pixel-wise velocity fields that capture the kinematic evolution of CME substructures. The derived velocity fields were validated through comparison with independently tracked height–time profiles. We found good agreement between the optical flow-derived velocities and the height–time measurements, confirming that the method reliably captures the kinematic evolution of CME substructures. Using this framework, we investigated the internal velocity dispersion between and within CME substructures. Our results reveal substantial spatial and temporal variability in CME internal dynamics, particularly in the middle corona \citep{West2023}, where both the front and core can undergo impulsive acceleration.

We found that the velocity evolution of CME substructures varies significantly with position angle (PA). Comparative analysis at different PAs revealed notable azimuthal variability in front–core dynamics. In the 2025 July 16 ASPIICS CME, all PAs exhibited a nearly synchronous peak impulsive acceleration in both the front and core (Figure~\ref{fig:aspiics20250716}). However, differences emerged during the decay phase: while some PAs maintained momentum coupling, others showed clear signs of phase lag and decoupling. For instance, at PA~283°, the front velocity dropped sharply and briefly rebounded after the core’s peak, producing transient out-of-phase behaviour. This suggests that such asynchrony arises during the post-impulsive acceleration phase, likely driven by internal momentum redistribution or differential expansion. Similar trends were observed in other events, such as the 2024 March 29 (Figure~\ref{fig:metis20240329}) and 2024 April 16 CMEs (Figure~\ref{fig:metis20240416}). In the latter, the core underwent stronger impulsive acceleration than the front, eventually out-speeding it and reducing its distance from the front, as reported in similar cases by \citet{Song2025}. Furthermore, as evident in the velocity-time profiles, the front and the core tend to decouple or propagate non-self-similarly after the impulsive acceleration phase. These findings reinforce the conclusion that CME substructures do not expand as rigid, self-similar flux rope shells \citep{Subramanian2014}, but instead exhibit complex, direction-dependent kinematics. Such asymmetries may arise from localised or internal reconnections along the flux rope \citep{Gibson2006}.

A key outcome of this study concerns the evolution of height dispersion between CME substructures. In most events that exhibited impulsive acceleration, whether in the front, core, or both, we observed that the height offset between the front and core began increasing only after the impulsive phase concluded. This was clearly seen in the 2025 July 16 ASPIICS and 2024 April 16 METIS events, and was also observed in the 2024 March 29 METIS event at PA~151°. These observations suggest that impulsive acceleration likely plays a central role in initiating internal restructuring and front–core separation. Interestingly, even in the 2024 March 29 event at PA 136°, where no impulsive acceleration was observed, a gradual increase in dispersion still occurred. Across all cases, height dispersion onset typically occurred between 2.3 and 3.0~$R_\odot$, significantly higher than the 1.4–1.8~$R_\odot$ range reported by \citet{Majumdar2024} based on inner-corona observations. This discrepancy likely arises from differences in instrumental coverage and resolution. In particular, the K-Cor field of view used in \citet{Majumdar2024} is confined to 3~$R_\odot$ and a cadence of 2 minutes, and the CME leading edge could be tracked only up to $\sim$2.6~$R_\odot$, emphasising the early evolution of CMEs in the inner corona. In contrast, our analysis focuses on heights above $\sim$1.9~$R_\odot$ and follows the continued evolution of the front–core separation within the middle corona using higher cadence and improved spatial resolution. Future studies using ASPIICS Level~3 HDR composites, expected to provide coverage over $\sim$1.1–3~$R_\odot$, may help bridge this gap across these height ranges.

Our analysis of radial self-similar expansion, applied independently to the CME front and core via the $S_r$ coefficient \citep{Wood2016}, reveals a systematic asymmetry between the two substructures across all four events: the core remains close to self-similar or super-self-similar ($S_r \gtrsim 1$) throughout the tracked height range, whereas the front is predominantly non-self-similar and contracting in the radial direction and, in three of the four events, shows a recovery toward self-similar values only at larger heights following an episode of strong contraction closer to the Sun. Hence, the core tends to expand much faster than the front at lower heights. We interpret this front--core asymmetry as arising from the front being constantly compressed between the outward push of the expanding core behind it and the ram pressure of the ambient corona ahead of it, while the core remains comparatively shielded from the ram pressure by the front. A quantitative test of this picture, however, requires independent density or ram-pressure diagnostics that lie beyond the scope of this work. We also find a positive correlation, though statistically non-significant given our sample of four events, between the front's time-averaged velocity and its degree of self-similarity, suggesting that faster fronts or more energetic CMEs tend to expand more self-similarly by resisting the ambient ram pressure. Independently, the angular width of three of the four CMEs increases rapidly at lower heights, and the expansion slows down at larger heights, consistent with the non-self-similar angular expansion of CMEs in the inner corona. The heights at which the angular width becomes stable vary across events (2.9--4.5~$R_\odot$). Taken together, these results indicate that the CME front and core evolve as dynamically distinct entities during propagation.

Our analysis of intra-structural dispersion further indicates that different regions within a single CME substructure do not evolve uniformly (Figure~\ref{fig:velocity_dispersion_regions}). The core showed a relatively narrow velocity distribution, consistent with uniform propagation, whereas the front displayed a wider velocity spread. In Section \ref{sec:self-similarity}, we found that the front frequently undergoes episodes of contraction (negative $S_r$) rather than self-similar expansion, indicating that its wider velocity spread reflects genuine non-uniform, and at times contracting, internal dynamics rather than a simple proportional scaling of velocity with distance. These internal asymmetries highlight that CME fronts are not kinematically homogeneous but evolve under localised forces and geometric influences, and can be effectively resolved using optical flow methods that capture motion across the structure.

From a technical perspective, we emphasized the importance of selecting an optimal differencing interval when preparing image sequences for optical flow. A short interval introduces dark trailing artifacts, while a long interval smears features and suppresses detail. Furthermore, to address flickering in ASPIICS 10-second exposure data, we formulated a Gaussian-tapered Fourier (GTF) filter that significantly improved velocity estimation stability. However, a more optimized method is needed to effectively mitigate flickering in the inner corona (or the 1-second exposure channel).

Also, it should be noted that our study is based on four CME events; hence, a more detailed statistical analysis is needed to reach a conclusion. Moreover, our method measured POS velocities, which introduces projection effects. To estimate the true velocities, we can incorporate the polarization-ratio technique \citep{Moran2004}, which provides 3D information about how far a given feature is from the sky plane and combine it with the optical flow velocity map to derive the true velocities. In future work, we plan to develop an automated CME detection approach based on frame-by-frame velocity distribution statistics, such as changes in histogram shape, spread, and high-velocity incidence. This method builds on our observation that CME passage is marked by a sharp rise in high-velocity magnitudes, followed by a return to background levels, offering a physics-driven alternative to traditional image-based techniques.

In conclusion, dense optical flow tracking reveals that CMEs exhibit internal velocity dispersion that varies with both height and azimuth, which is best captured using high-resolution, high-cadence coronagraph observations. Furthermore, recent and upcoming missions such as PUNCH \citep{DeForest2025} and Vigil \citep{West2025} will enable investigations of internal CME kinematics at greater heliospheric distances and from the L5 vantage point, respectively, advancing our ability to model and forecast space weather impacts. Importantly, such internal kinematic complexity has direct implications for improving these forecasts, especially for geo-effective CMEs \citep{Temmer2021}. Models like the Drag-Based Model (DBM) typically rely on the bulk CME front speed to predict arrival times at Earth \citep{Vrsnak2013, Dumbovic2021}. However, our findings suggest that significant velocity gradients can exist within the CME, with the core and the front potentially decoupling in the middle corona. Incorporating internal kinematic information into such models could enhance their predictive power.

The code used for image preprocessing and computing dense optical flow map is publicly available at \href{https://github.com/pritamd9818/DOFCAT.git}{DOFCAT GitHub Repository}.

\begin{acknowledgments}
We thank the anonymous reviewer for the insightful
comments and suggestions, which have greatly helped in improving the manuscript. ASPIICS data are courtesy of Proba-3/ASPIICS. Proba-3 is a technology demonstration mission of the European Space Agency (ESA) and a Mission of Opportunity in the ESA Science Programme. Solar Orbiter is a space mission of international collaboration between ESA and NASA, operated by ESA. Metis was built and operated with funding from the Italian Space Agency (ASI), under contracts to the National Institute of Astrophysics (INAF) and industrial partners. Metis was built with hardware contributions from Germany (Bundesministerium für Wirtschaft und Energie through DLR), from the Czech Republic (PRODEX), and from ESA. The EUI instrument was built by CSL, IAS, MPS, MSSL/UCL, PMOD/WRC, ROB, LCF/IO with funding from the Belgian Federal Science Policy Office (BELSPO/PRODEX PEA C4000134088); the Centre National d’Etudes Spatiales (CNES); the UK Space Agency (UKSA); the Bundesministerium für Wirtschaft und Energie (BMWi) through the Deutsches Zentrum für Luft- und Raumfahrt (DLR); and the Swiss Space Office (SSO). P.D. and V.P. would like to thank Pallavi Rajeev for the initial demonstration of the optical flow algorithm on SOHO/LASCO C2 data. P.D. would also like to thank Laurent Dolla and Andrei Zhukov for discussions on mitigating brightness flickering in ASPIICS images. We thank Dipankar Banerjee for insightful discussions regarding this work. P.D. acknowledges financial support from the Innovation in Science Pursuit for Inspired Research (INSPIRE) fellowship award DST/INSPIRE Fellowship/2021/IF210727.

\end{acknowledgments}

\bibliography{references}
\bibliographystyle{aasjournalv7}

\end{document}